\documentclass[prc,preprint]{revtex4}
\usepackage{graphicx}
\usepackage{subfigure}
\usepackage{amsmath,amssymb}
\usepackage{mathrsfs}
\def\dis{distribution}

\begin{document}
\title{Multi-minijet Contribution to Hadronic Spectra and Correlations in Pb-Pb Collisions at 2.76 TeV and beyond}
\author{Rudolph C. Hwa$^1$ and Lilin Zhu$^{2}$}
\affiliation
{$^1$Institute of Theoretical Science and Department of
Physics\\ University of Oregon, Eugene, OR 97403-5203, USA\\
$^2$Department of  Physics, Sichuan
University, Chengdu  610064, P.\ R.\ China}
\date{\today}

\begin{abstract}
In heavy-ion collisions at very high energy the density of produced jets can be so high that the possibility of hadrons produced by recombination of shower partons in overlapping minijets may become important. We study such multi-minijet contribution to the hadron spectra and to dihadron correlation in Pb-Pb collisions at 2.76 TeV at LHC. We adjust the parameter controlling the momentum degradation of semihard partons by fitting the charged-particle distribution up to $p_T\sim 16$ GeV/c. The relative magnitudes of different identified hadrons and of various partonic components are fixed by the recombination formalism. We find that the coalescence of shower patons from adjacent miniijets can be as much as from single jets for meson production, and even more so for proton, but never dominant over other components. In 3-shower-parton recombination the ratio of 2-jet to 1-jet contributions increases with collision energy; its maximum can exceed 2 at 5.5 TeV. Two-hadron correlation exhibits a broad peak on transverse rapidities, confirming that minijets play a central role at low $p_T$.

\pacs: {25.75.Dw, 25.75.Gz}
\end{abstract}

\maketitle
\section{Introduction}

In heavy-ion collisions above 2 TeV the density of minijets produced by semihard scatterings of partons can be so high that conventional treatment of such collisions may be inadequate.  What is conventional at lower energy (0.2 TeV) is hydrodynamical description for transverse momentum $p_T < 2$ GeV/c \cite{kh} and jet fragmentation at $p_T > 6$ GeV/c \cite{gv}.  In the intermediate region neither approach is reliable.  The unconventional treatment based on quark recombination has shown some success in filling the gap \cite{hy, gk, fm}.  Conceptually, one can imagine that at extremely high energies, higher than what is available at the CERN Large Hadron Collider (LHC), heavy-ion collisions can be so explosive with a preponderance of minijets that hydrodynamics would be invalid even at low $p_T$ on the one hand, and absence of jet-jet interaction would be unreasonable at moderate $p_T$ on the other.  The question is what the reality is at LHC where charged particle distribution is already known up to $p_T \sim 18$ GeV/c \cite{ka}.  An answer cannot be given without some model-dependent analysis.  Our aim in this paper is to find the degree of importance of multijet recombination in the formation of hadrons.

Recently, the spectra of identified hadrons in Pb-Pb collisions at LHC for $p_T < 5$ GeV/c have been studied \cite{mf,hz}; it is found that the recombination of thermal and shower partons is important down to $p_T \sim 1$ GeV/c.  It means that there is an abundance of minijet produced at $\sqrt{s_{NN}} = 2.76$ GeV.  The basic parameters ($T$, inverse slope of thermal partons, and $\kappa$, a measure of the momentum degradation of semihard partons) are determined by fitting the data on $\pi, K, p$ and $\Lambda$ distributions simultaneously.  With those parameters at hand, we can then extend the study to higher $p_T$ with sufficient constraint to confront the  data on charged hadrons.

 Low-energy jets with transverse energy $E_T\ ^<_\sim\ 10$ GeV are not the sort of objects that can be identified by jet-finding algorithms, such as HIJA \cite{sb}, for heavy-ion collisions because they merge into the backgrounds from high-multiplicity underlying events in search for jets with $E_T > 50$ GeV \cite{cms}.  Separation of neighboring minijets is not only unfeasible, but probably not meaningful.  A  discussion of jets usually starts with a definition of what the minimum energy in a jet cone $(E_T^{\rm cone})$ is after background subtraction, and what the radius of the jet cone $(R)$ is.  It is based on the concept that particles in a jet are correlated with the initiating hard parton, and can be isolated from the background by an effective algorithm.  The task for accomplishing that becomes more and more difficult as the jet energy is reduced because of the fluctuation of the background.  From the point of view of $p_T$ distribution that is averaged over all pseudorapidity $\eta$ and azimuthal angle $\phi$, at least in a midrapidity interval, say $|\eta| < 1$, and over all events, the background is the dominant part that is exponential.  Small deviation from that part at a slightly larger $p_T$ reveals the effect of semihard scatterings that cannot be easily identified on the event-by-event basis as well-defined jets in terms of $E_T$ and $R$.  Nevertheless, we need to focus on them in order to calculate their effect on the $p_T$ distribution.  For that purpose we use the term minijet without being precise about $E_T$ and $R$, since we shall not examine the event structure in the $\eta$-$\phi$ plot.  We do consider semihard partons with transverse momenta $k > 3$ GeV/c and the shower partons that they generate after emerging from the medium surface.  Each such semihard parton and the cluster of associated shower partons will be referred to as a minijet.

In heavy-ion collisions there are various theoretical issues related to minijets that have not yet evolved to a mature subject with general acceptance. The medium effects on semihard partons are important but hard to make precise, and the hadronization process is still controversial. The shower partons not only depend on the degree of momentum degradation in the medium, but also have various channels of hadronization, such as through recombination with thermal partons on the one hand and with other shower partons on the other. At LHC the  high density of jets creates the possibility of shower partons from different jets
overlapping in common spatial proximity so that their coalescence cannot be ignored. The study of multi-minijet contribution to the hadronic $p_T$ distribution and to two-particle correlation is the main concern of this paper in addition to the usual components involving single jets. We shall examine all possible components so as to exhibit the relative importance of each up to $p_T\sim 16$ GeV/c. From $\sqrt s=2.76$ TeV with known charged-particle spectra, we extrapolate to 5.5 TeV to show how the multi-minijet contributions depend on collision energy.

Although we can calculate the minijet contribution to the hadronic spectra, it is difficult to conclude from the observed $p_T$ spectra that minijets are necessarily existent. To make that conclusion cogent, two-particle correlation exhibited in terms of transverse rapidities has been used experimentally to show the existence of a broad peak \cite{lr}. We shall analyze our calculated results in those variables and show general agreement with the data. Thus there is common ground in recognizing the important role that minijets play in heavy-ion collisions. However, our approach differs from the experimental approach in that we provide the partonic basis of the observed phenomenon.
At low $p_T$ where correlation is strongest, thermal partons play an important role despite the fact that they are by themselves uncorrelated.  That is because two shower partons from one jet are correlated by momentum constraints, so their separate recombination with different thermal partons results in hadronic correlation.  Thus $p_T$ correlation among particles that are not too far apart in pseudorapidity $\eta$ and azimuthal angle $\phi$ is a fertile ground to find the footprints of recombination.

Our study here is limited to midrapidity in central collisions and all our formulas are averaged over the azimuthal angles. As will become self-evident, the theoretical description that includes all components of hadronization (e.g., 7 components for proton production) is sufficiently complicated that it becomes essential to build first a clear and solid foundation for the physical processes without involving the azimuthal complexities. We shall use schematic diagrams to help the visualization of the various processes.

In the next two sections we describe the multijet recombination processes for pions and protons. Two-particle correlations are discussed in Sec.\ IV. The parameters we use to do detailed calculations are given in Sec.\ V with results on single-particle spectra shown in Sec.\ VI. Extension to 5.5 TeV is made in Sec.\ VII. The results on two-particle correlation are presented in Sec.\ VIII. Conclusion is then given in the final section.

\section{Two-jet Recombination for Pion Production}

All basic equations for the inclusive $p_T$ distributions of hadrons ($\pi, K, p, \Lambda$) produced in Pb-Pb collisions at LHC have already been given explicitly in Ref.\ \cite{hz}, whose equations will be referred to hereafter with the prefix I.
The definition of all the quantities undefined in this paper can be found in \cite{hz}.
We shall not repeat them here.  Instead, we show schematic diagrams that can clearly depict the processes involved.  They include both the types of processes already considered in \cite{hz} and new ones, such as two-jet recombination, for which relevant equations will be given.

 \begin{figure}[tbph]
\centering
\vspace*{-8cm}
\includegraphics[width=0.8\textwidth]{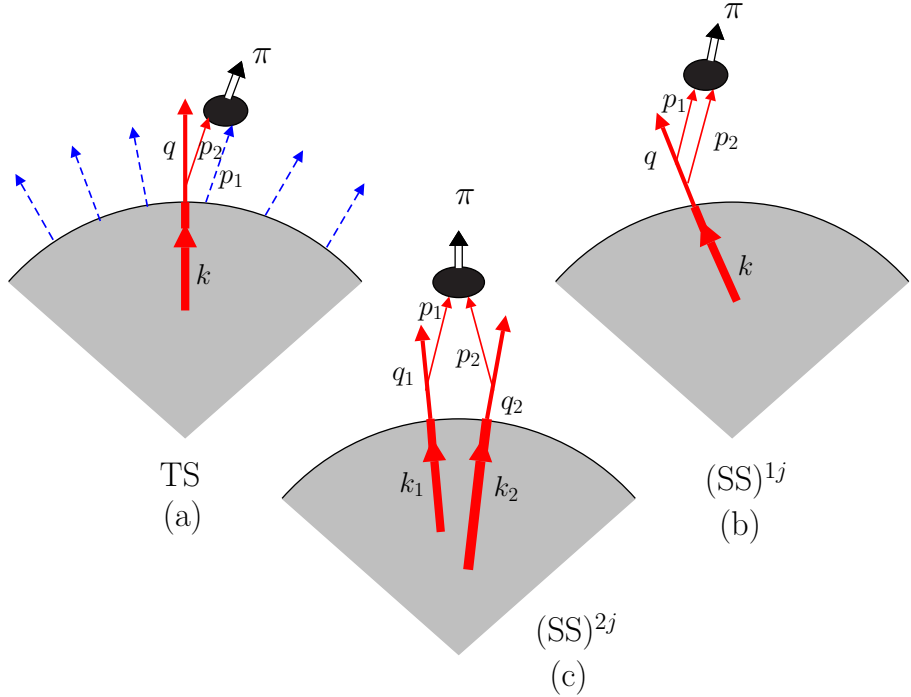}
\vspace*{-.5cm}
\caption{(Color online) Schematic diagrams for parton recombination of (a) TS, (b) SS in one jet, and (c) SS in two jets. Thick (red) lines represent partons in medium, thin (red) lines partons out of medium, thinnest (red) lines shower partons, and dashed (blue) lines thermal partons.}
\end{figure}

In Fig.\ 1 we show the diagrams in the transverse plane for the recombination of (a) thermal T and shower S partons, (b) SS in one jet, and (c) SS in two jets, which will be abbreviated by TS, (SS)$^{1j}$ and (SS)$^{2j}$, respectively.  In the notation of Eq.\ I-(35), $k$ is the momentum of the semihard parton at creation, and $q$ is the momentum at the medium surface.  The thick red vectors have the dual role of representing the jet momentum in the medium and the degradation effect described by $G(k,q, \xi)$ in the same equation \cite{hz,hy1}.  The thinner red lines outside the medium are the semihard partons $q_j$, which can emit shower partons represented by the thinnest red lines denoted by $p_j$.  The blue dashed arrows are thermal partons.  Recombination is presented by a large black blob with the outgoing open arrow depicting the produced pion.  The lengths and angles of the vectors are not drawn to scale due to the limitation in presenting the figures clearly, and should not be taken literally.

Figure 1(a) is described by I-(14), and Fig.\ 1(b) by I-(7), (18).  For Fig.\ 1(c) to occur, the two vectors $\vec k_1$ and $\vec k_2$ should be nearby and approximately parallel so that there is an appreciable overlap of the jet cones.  Considering only partons that are near $\eta=0$, and exhibiting the $\phi$ dependence, we have for the two-jet contribution to the pion spectrum
\begin{eqnarray}
p^0{dN_{\pi}^{2j}\over dp_Td\phi} = \int \prod_{j=1}^2 \left[{dp_j\over p_j}d\phi_jd\xi_jP(\xi_j,\phi_j,b){\cal S}(p_j,\xi_j)\right]  {\bf
R}_\Gamma^\pi (p_1,\phi_1,p_2,\phi_2,p_T,\phi),   \label{2.1}
\end{eqnarray}
where ${\cal S} (p,\xi)$ is the integrated shower parton distribution defined in I-(6) in terms of the semihard parton distribution $F_i(q,\xi)$ at the surface and the shower distribution $S_i(p/q)$ in the $i$-jet, integrated over $q$ and summed over $i$.  The variable $\xi$ is a measure of the dynamical path length in the medium created by the heavy-ion collision at impact parameter $b$ \cite{hy1}.  $P(\xi,\phi,b)$ is the probability for $\xi$ to occur for a path at angle $\phi$ initiated at ($x_0, y_0$), weighted by the nuclear overlap function, and integrated over all ($x_0, y_0$).  The quantity in the square brackets in Eq.\ (\ref{2.1}) is the probability of having shower parton $p_j$ in Fig.\ 1(c).  $R_{\Gamma}^{\pi}$ is the recombination function (RF) characterized by $\Gamma$ that summarizes all other dependencies besides $p_j$ and $\phi_j$, such as the spatial separation of the shower partons.  It is important to recognize that the product ($j=1, 2$) in Eq.\ (\ref{2.1}) implies the creation and degradation of two independent semihard partons, but whose shower partons must be in close proximity outside the medium if they are to coalesce to form the pion.  Thus only at high energy (as at LHC) is the jet density high enough for the 2-j recombination to occur.

For the RF, ${\bf R}_\Gamma^\pi(p_1,\phi_1,p_2,\phi_2,p_T,\phi)$, the coalescence process clearly cannot take place if $\phi_1$ and
$\phi_2$ are not nearly equal, since non-parallel partons have large relative momentum transverse to $\vec p_1 + \vec p_2$.  Large relative
longitudinal momentum parallel to $\vec p_1 + \vec p_2$  is permitted in the parton model, since the momentum fraction of a parton in a hadron can
vary from 0 to 1.  Relative momentum transverse to that is limited by the confinement restriction that it should not exceed the binding energy of the
constituents.  One may consider a Gaussian distribution in $|\phi_1 - \phi_2|$ with an appropriate width.  However, since $\phi_1$ and $\phi_2$ are
integrated over in Eq.\ (\ref{2.1}), it is simpler to adopt a factorizable form that requires the partons to be parallel but with a suitable
normalization factor $\Gamma$ that we can estimate, i.e.,
\begin{eqnarray}
{\bf R}_\Gamma^\pi(p_1,\phi_1,p_2,\phi_2,p_T,\phi) = \Gamma\delta(\phi_1-\phi_2)\delta\left({\phi_1+\phi_2\over 2} - \phi\right) R^\pi(p_1,p_2,p_T),
\label{2.2}
\end{eqnarray}
where $\Gamma$ is the probability that two parallel partons can recombine.  Since the partons are emitted from the medium at early times, we may
consider the emitting system as being a thin almond-shaped overlap region viewed from its side in the same transverse plane at midrapidity as where
the pion is detected.  For centrality $c<0.05$ the almond is almost circular.  The partons at $\phi_i$ are parallel, but can be emitted at any
distance from the center of the circle.  Looking at the emitting source edgewise, it is essentially a one-dimensional system of width approximately 10
fm, which is slightly less than $2R_A$ since high-density partons are not likely to be emitted tangentially from the edges.  The two parallel partons
should be separated by a distance not greater than the diameter of a hadron ($\sim 2$ fm), given that the jets have some width.  Thus our estimate for
$\Gamma$ is the ratio $\sim 2/10$.  We do not see that any more elaborate analysis of the coalescence process can provide a more transparent
description of ${\bf R}_\Gamma^\pi$.

Applying Eq.\ (\ref{2.2}) to (\ref{2.1}), we obtain
\begin{eqnarray}
{dN_\pi^{2j}\over p_Tdp_T} = {\Gamma\over p_T^2} \int {dp_1\over p_1}{dp_2\over p_2} \int \prod_{\alpha=1}^2 \left[{dq_{\alpha}\over q_\alpha} \sum_i \bar F_i(q_\alpha, \kappa)S_i(p_\alpha/q_\alpha)\right]R(p_1,p_2,p_T),     \label{2.3}
\end{eqnarray}
where the integral over $\xi_j$ in (\ref{2.1}) has been replaced by the average distribution $\bar F_i(q, \kappa)$, discussed in \cite{hz}.  Further consideration of the parameter $\kappa$ will be explained below, but for now we give the original definition
\begin{eqnarray}
1/ \kappa = q/k     \label{2.4}
\end{eqnarray}
which is the fraction of the semihard parton's momentum $k$ that is not lost to the medium as it emerges with momentum $q$.  All parts of Eq.\ (\ref{2.3}) are known from \cite{hz}, so it is straightforward to calculate the 2-j contribution to pion production.

It should be noted that theoretically it is possible for a very hard parton to split into two softer partons which can separately form two jets that are close-by and may even overlap. Such a QCD process would have only one $F_i$ function at a much higher parton momentum. However, since the parton distribution $f_i(k)$ at creation decreases rapidly with increasing $k$, the production of such hard partons is suppressed compared to the semihard partons that we consider, as our results will show the dominance of minijets. For that reason such processes are not included in our
calculation of $dN_\pi^{2j}/p_Tdp_T$.

 \begin{figure}[tbph]
\centering
\vspace*{-3cm}
\includegraphics[width=0.8\textwidth]{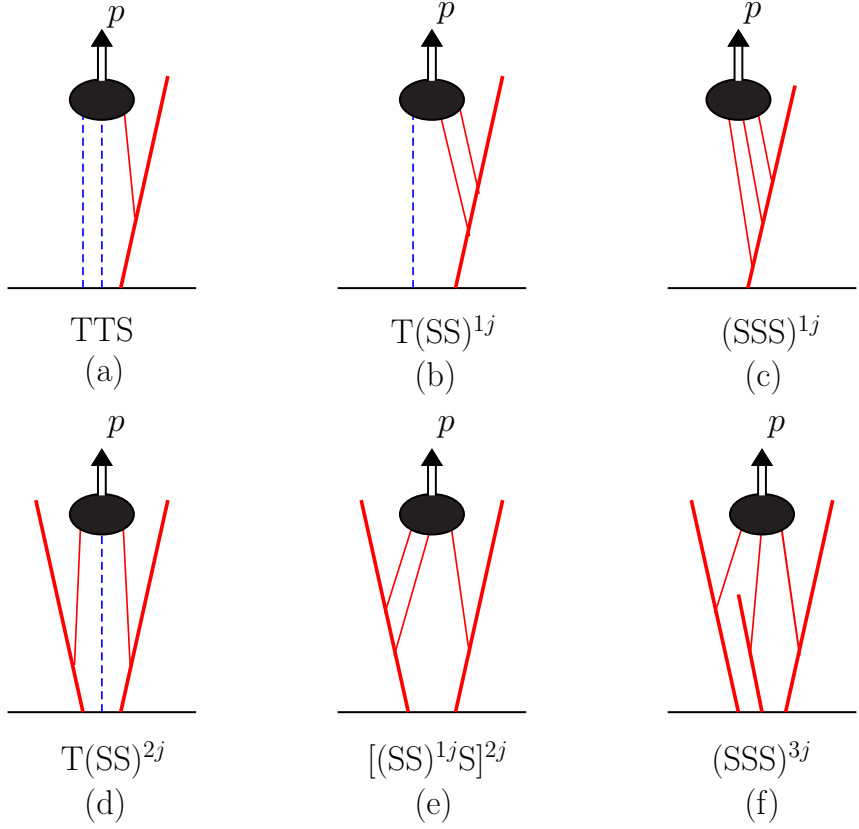}
\vspace*{-5cm}
\caption{(Color online) Diagrams for proton production by recombination of partons with same line-types as in Fig.\ 1.}
\end{figure}

\section {Multijet Recombination for Proton Production}

There are many components of thermal and shower recombination that contribute to the production of protons.  Symbolically, they can be expressed as in I-(19)
\begin{eqnarray}
F_{qqq} = {\cal TTT + TTS + T(SS)}^{1j} + {\cal (SSS)}^{1j} + {\cal T(SS)}^{2j} +[{\cal  (SS)}^{1j}{\cal S]}^{2j} + {\cal (SSS)}^{3j}   \label{3.1}
\end{eqnarray}
Except for the first term that does not involve any S, the other six terms are depicted by the six figures in Fig.\ 2, respectively.  The equations for TTS, T(SS)$^{1j},$ and (SSS)$^{1j}$, corresponding to Fig.\ 2(a), (b) and (c), are given in I-(A9), (A10) and (A11), respectively.  For the next three processes they are the following
\begin{eqnarray}
{dN_p^{{T(SS)}^{2j}}\over p_Tdp_T} = {g_{\rm st}^pN_pC\Gamma\over m_T^p p_T^{2\alpha+\beta+3}} \int_0^{p_T} dp_1 \int_0^{p_T-p_1}
dp_2(p_1p_2)^{\alpha}(p_T-p_1-p_2)^{\beta} \nonumber \\
 \times {1\over 3} \sum_{(jk\ell)} p_je^{-p_j/T}{\cal S}^q(p_k,\kappa){\cal S}^q(p_\ell,\kappa).   \label{3.2}
\end{eqnarray}
\begin{eqnarray}
{dN_p^{[({SS)}^{1j}{S}]^{2j}}\over p_Tdp_T} = {g_{\rm st}^pN_p\Gamma\over m_T^pp_T^{2\alpha+\beta+3}} \int_0^{p_T} dp_1 \int_0^{p_T-p_1}
dp_2(p_1p_2)^{\alpha}(p_T-p_1-p_2)^{\beta} \nonumber \\
\times {1\over 3} \sum_{(jk\ell)} {\cal S}^{qq}(p_j,p_k,\kappa){\cal S}^q(p_\ell,\kappa).   \label{3.3}
\end{eqnarray}
\begin{eqnarray}
{dN_p^{({SSS})^{3j}}\over p_Tdp_T} = {g_{\rm st}^pN_p\Gamma^2\over m_T^pp_T^{2\alpha+\beta+3}} \int_0^{p_T} dp_1 \int_0^{p_T-p_1}
dp_2(p_1p_2)^{\alpha}(p_T-p_1-p_2)^{\beta} \prod_{j=1}^3 {\cal S}^q(p_j,\kappa),  \label{3.4}
\end{eqnarray}
where $\sum_{(jk\ell)}$ denotes cyclic permutation of $(j,k,\ell)$ over (1,2,3) with $p_3 = p_T-p_1-p_2$.
The exponents $\alpha=1.75$ and $\beta=1.05$ are from the proton RF \cite{hy,hz}.
The quantities ${\cal S}^q$ and ${\cal S}^{qq}$ are defined in I-(A4) and (A12) for one quark and two quarks in a gluon jet.  The sum over all semihard partons is approximated by $\sigma$ times gluon contribution only with $\sigma = 1.2$ signifying that all other quark jets are regarded as contributing $\sim 20\%$ more to the gluon jet.  This approximation is based on concrete calculations of certain components in which we compare the $u$ quark contribution to that of the gluon.  We have found that using $\sigma = 1.2$ as an average multiplicative factor on $\bar F_q(q,\kappa)$ is a reasonable approximation of $\sum_i\bar F_i(q,\kappa)$, which, if exhibited in detail, would be overwhelmingly complicated without rendering significant elucidation or accuracy to justify showing them.

The results of our calculation of all six terms shown in Fig.\ 2 will be exhibited below.  The same procedure can be applied to the determination of hyperon spectra, but will not be pursued here.

\section {Two-particle Correlation}

A minijet refers to a cluster of particles generated by a semihard parton that emerges from the medium. Those particles must be correlated since each
of those hadrons in the cluster must consist of at least one shower parton generated by the same semihard parton. The correlation among those multiple
shower partons results in correlation among the hadrons, even though the hadronization process may involve thermal partons. High-$p_T$ jets are routinely studied by jet algorithms, which are, however, ineffective for minijets.  Since we do not pursue the issue of angular correlation in this paper, that having been done already in Ref.\ \cite{ch}, we focus here on the correlation in the $p_T$ variables.  The $\eta$ and $\phi$ variables of the hadrons under consideration are in close proximity, since the correlations are either within one jet or in overlapping adjacent jets.  Only averages over $\eta$ and $\phi$ are calculated.

Let us define the Pearson's covariance, as used in \cite{lr,md,tk,ht,tr},
\begin{eqnarray}
P_2(1,2)={C_2(1,2)\over [\rho_1(1)\rho_1(2)]^{1/2}} \ , \quad \qquad
C_2(1,2)=\rho_2(1,2)-\rho_1(1)\rho_1(2) \ ,   \label{4.1}
\end{eqnarray}
where
\begin{eqnarray}
\rho_1(1)={dN_{h_1}\over p_1dp_1}\ ,  \qquad \quad
\rho_2(1,2)={dN_{h_1h_2}\over p_1dp_1p_2dp_2} \ .   \label{4.2}
\end{eqnarray}
We shall calculate $P_2(p_{1_T},p_{2_T})$ for $\pi\pi$ and $pp$ correlations, but discuss mainly the former.  We shall use $p_t$ and $p_a$ (instead of $p_1$ and $p_2$) to denote the momenta of the two hadrons ($t$ for trigger, $a$ for associated particle, although the two particles are treated on equal footing in $\rho_2(1,2)$) in order to avoid notational confusion with the parton momenta $p_i$ already used in Secs. II and III and in Fig.\ 1.

The background subtraction in Eq.\ (\ref{4.1}) is automatically taken into account if we consider only the non-factorizable terms in $\rho_2(1,2)$.  There are six such non-factorizable terms, which are shown schematically in Fig.\ 3.  They are denoted as:  (a) (TS)(TS), (b) (TS)(SS)$^{1j}$, (c) (TS)(SS)$^{2j}$, (d) (SS)$^{1j}$(SS)$^{1j}$, (e) (SS)$^{1j}$(SS)$^{2j}$, and (f) (SS)$^{2j}$(SS)$^{2j}$.  The corresponding equations for the correlated distributions are given in Appendix A.

 \begin{figure}[tbph]
\centering
\vspace*{-1cm}
\includegraphics[width=0.8\textwidth]{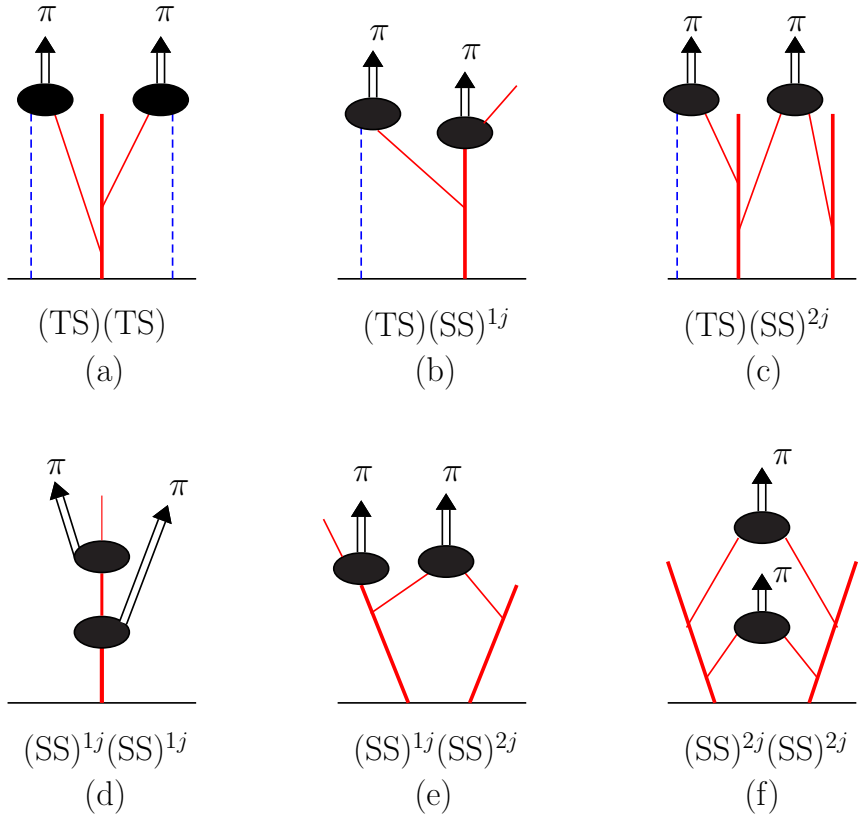}
\vspace*{-5cm}
\caption{(Color online) Diagrams for two-pion correlation with same line-types as in Fig.\ 1.}
\end{figure}

Consider first (TS)(TS) correlation. The two shower partons must  come from the same semihard parton in order for them to be correlated. ${\cal
S}^{qq}(p_1,p_2,\kappa)$ defined in Eq.\ I-(A12) describes the distribution of those two shower partons $p_1$ and $p_2$, integrated over all $q\ge p_1+p_2$.
Note that their correlation is due primarily to momentum constraints, although charge correlation (or more generally quark-type correlation) cannot be
excluded. For simplicity, we ignore the constraints arising from quark types, since their recombination with other uncorrelated partons (specially
thermal ones) partially neutralizes the effect. The parton correlation is transmitted to the two pions by the coalescence of the shower partons with
the thermal partons.

In (TS)(SS)$^{1j}$ the correlation arises from the shower parton in (TS) being emitted by the same semihard parton that fragments to the other pion.  It is obvious that since all the diagrams in Fig.\ 3 are connected, the two pions produced are not factorizable, and thus correlated.  The equations in Appendix A constitute the six terms of $C_2(1,2)$.    The demominator of $P_2(1,2)$ in Eq.\ (\ref{4.1}) must include all the factorizable terms also, involving not only (TT)$_i$, but also (TS)$_1$(TS)$_2$, etc., where the two shower partons are generated by two independent semihard partons.

Since it has been shown in \cite{hz} that the proton spectrum in the region $p_T < 5$ GeV/c is dominated by TTS recombination, and that the $p/{\pi}$ ratio decreases with increasing $p_T$ above the peak at around $p_T \approx 3$ GeV/c, to study the two-particle correlation between protons it is sufficient to investigate only the term (TTS)(TTS).  The diagram for that is shown in Fig.\ 4 and the equation for it is given in Appendix A.

 \begin{figure}[tbph]
\centering
\vspace*{-3cm}
\includegraphics[width=.9\textwidth]{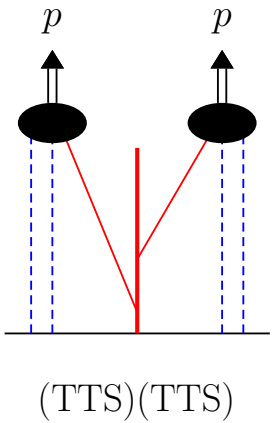}
\vspace*{-12cm}
\caption{(Color online) Diagram for two-proton correlation with TTS component only for each proton.}
\end{figure}

\section {Parameters}

There are two parameters adjusted to fit the ALICE data on $\pi, K, p, \Lambda$ spectra for $p_T < 5$ GeV/c \cite{mf}.  They are the inverse slope $T$ of the thermal parton distributions and the parameter $\kappa$ in Eq.\ (\ref{2.4}).  Their values are \cite{hz}
\begin{eqnarray}
T = 0.38 \ {\rm GeV}, \qquad\qquad  \kappa = 2.6.     \label{5.1}
\end{eqnarray}
Using these parameters we can calculate the spectra at higher $p_T$ including the multi-minijet contributions.  Although the data for the identified hadrons are not available for $p_T > 5$ GeV/c, the charged hadron spectrum has been measured up to 19 GeV/c.  Assuming that $\Sigma^+$ can be represented by $\Lambda$, the sum of what we can calculate, $\pi + K + p + \Lambda$, can be a good representation of the charged distribution, as we have demonstrated for $p_T < 5$ GeV/c in \cite{hz}.  Thus we proceed and carry out the calculation for the full $p_T$ range.  We discover, however, that the resultant charged spectrum is lower than the data at high $p_T$.  There is a reason for that, as we now describe.

In extending to high $p_T$ we have fixed $\kappa$ to be constant at the value given in Eq.\ (\ref{5.1}) that is determined in \cite{hz}.  It has the physical meaning that $\kappa^{-1}$ is the fraction of the semihard parton's momentum $k$ that is retained by the parton as it reaches the medium surface, since $k = \kappa q$.  Keeping $\kappa$ fixed implies that the effect of the medium on the energy loss is independent of the parton momentum.  That property is, however, not consistent with the data on the nuclear modification factor $R_{AA}$ at LHC.  For 0-5\% centrality in Pb-Pb collisions at 2.76 TeV ALICE has shown that $R_{AA}$ increases from 0.14 at $p_T = 6$ GeV/c to 0.35 at $p_T=19$ GeV/c \cite{ha}, as is expected from most energy-loss models.  It means that momentum degradation as a fraction of the initial momentum decreases with increasing parton momentum.  Thus it is necessary for us to consider a $q$-dependent $\kappa$, for which we use the form
\begin{eqnarray}
\kappa (q) = {\kappa_0\over 1 + \kappa_1q^2},     \label{5.2}
\end{eqnarray}
where $\kappa_0$ and $\kappa_1$ are constrained by $\kappa = 2.6$ at low $q \ ^<_\sim\ 10$ GeV/c.  We find by fitting the charged hadron spectrum that their suitable values are
\begin{eqnarray}
\kappa_0 = 3,  \qquad \quad \kappa_1 = 0.0018\ ({\rm GeV/c})^{-2}.     \label{5.3}
\end{eqnarray}
In the next section we show the results of our calculation based on these parameters and compare  them to the data.

Although the overall $p_T$ dependence is obtained by varying $\kappa_{0,1}$, the relative magnitudes of the various components are not independently adjustable.  Thus we learn about the nature of the hadronization processes that we cannot otherwise.  Moreover, the correlation properties can then be determined without further arbitrariness.

We note that since $\kappa(q)$ cannot be less than 1, there is an upper limit of $q^2$ for which Eq.\ (\ref{5.2}) can be used.  It corresponds to $q = 33$ GeV/c.  Any contribution from $q$ larger than that value would be very small so the invalidity of Eq.\ (\ref{5.2}) for very large $q$ has been ignored in our calculation.


 \begin{figure}[tbph]
\centering
\includegraphics[width=0.8\textwidth]{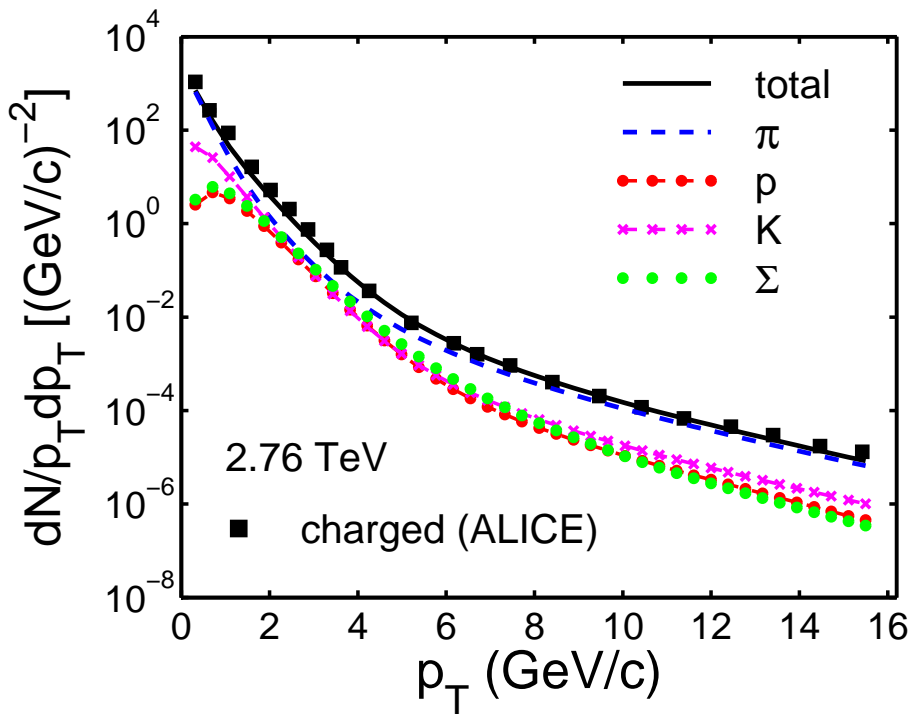}
\caption{(Color online) Charged particle distribution in $p_T$ calculated as the sum of four components for Pb-Pb collisions at 2.76 TeV. The data are from Ref.\ \cite{ka}.}
\end{figure}

\section {Results on Hadron Spectra}

We first show the charged hadron distribution which we identify as the sum of $\pi, K, p$ and $\Sigma$.  Each of the identified hadron spectra is calculated as in \cite{hz} but with all multijet contributions included here and with $p_T$ extended to 16 GeV/c.  The $\kappa$ parameter used is given in Eqs.\ (\ref{5.2}) and (\ref{5.3}).  The result is shown in Fig.\ 5, in which the black solid line is the sum that fits the data \cite{ka} very well.  The four hadronic components are shown separately by different lines.  Note that for $p_T >5$ GeV/c the pion spectrum dominates over the others, but for $p_T \sim 3$ GeV/c all four have nearly the same magnitude, resulting in the total to be noticeably larger than each individually in that region.

 \begin{figure}[tbph]
\centering
  \subfigure{
    \includegraphics[width=.47\textwidth]{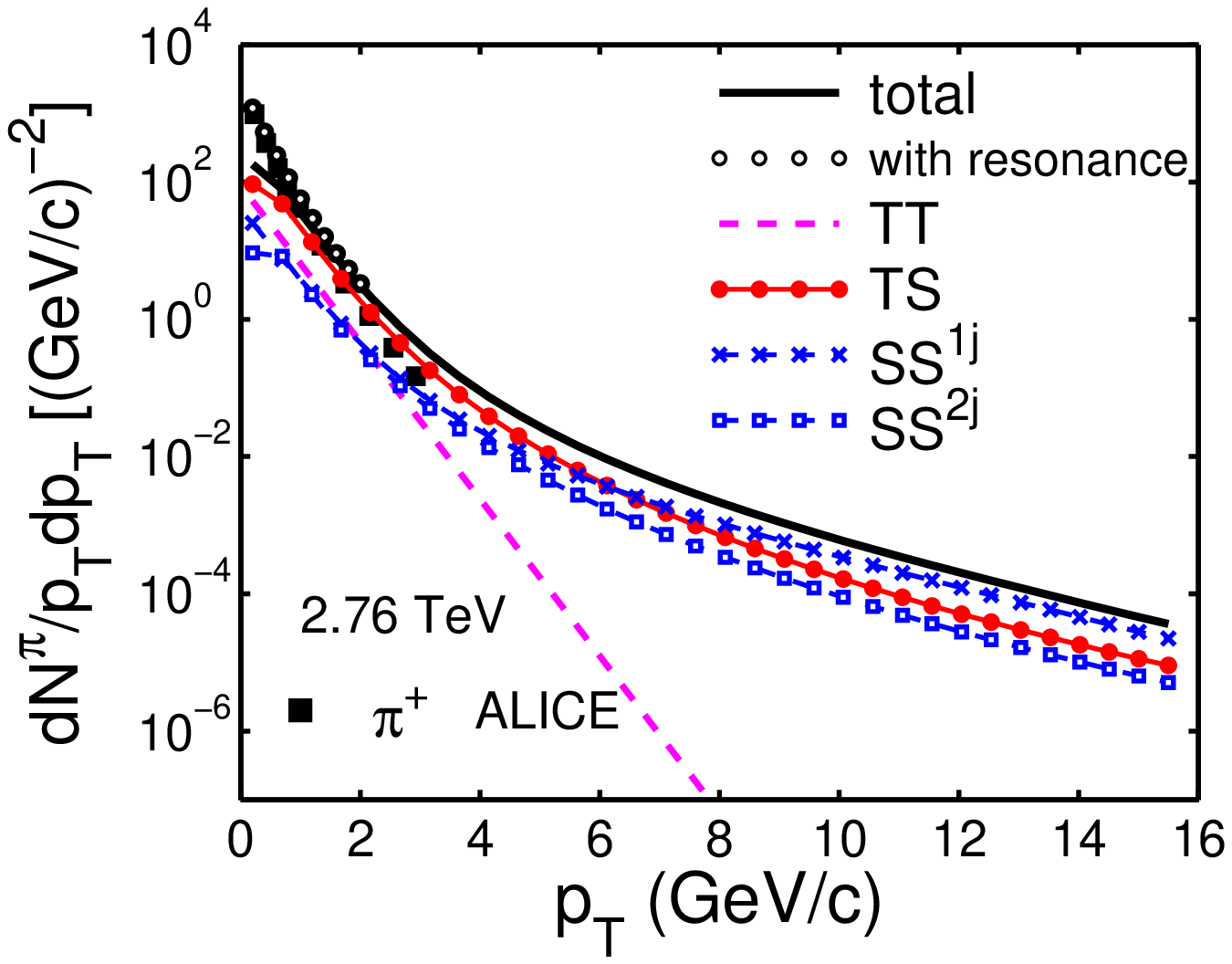}}
  \hspace{0.1in}
  \subfigure{
    \includegraphics[width=.47\textwidth]{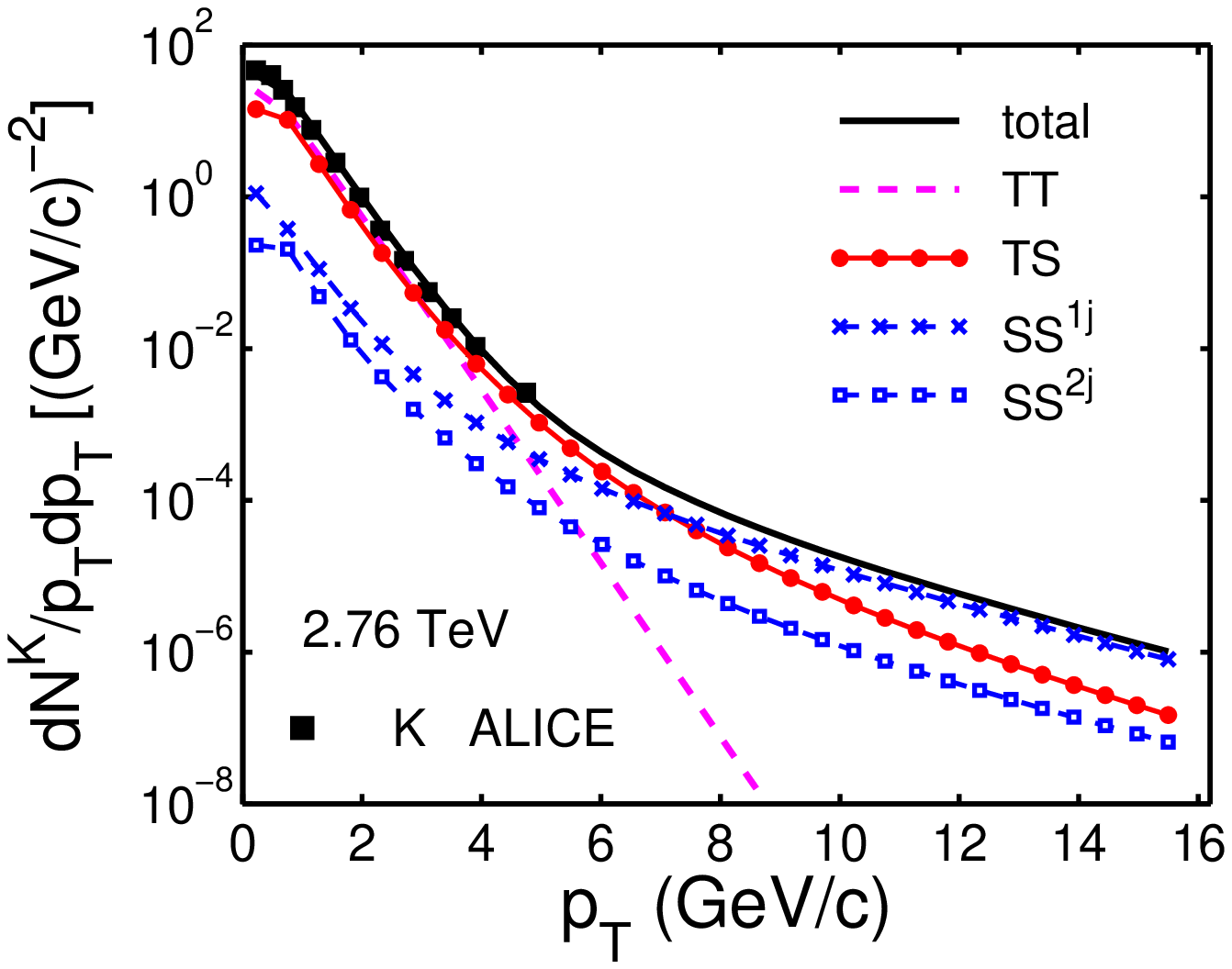}}\\[20pt]
\centering
  \subfigure{
    \includegraphics[width=.47\textwidth]{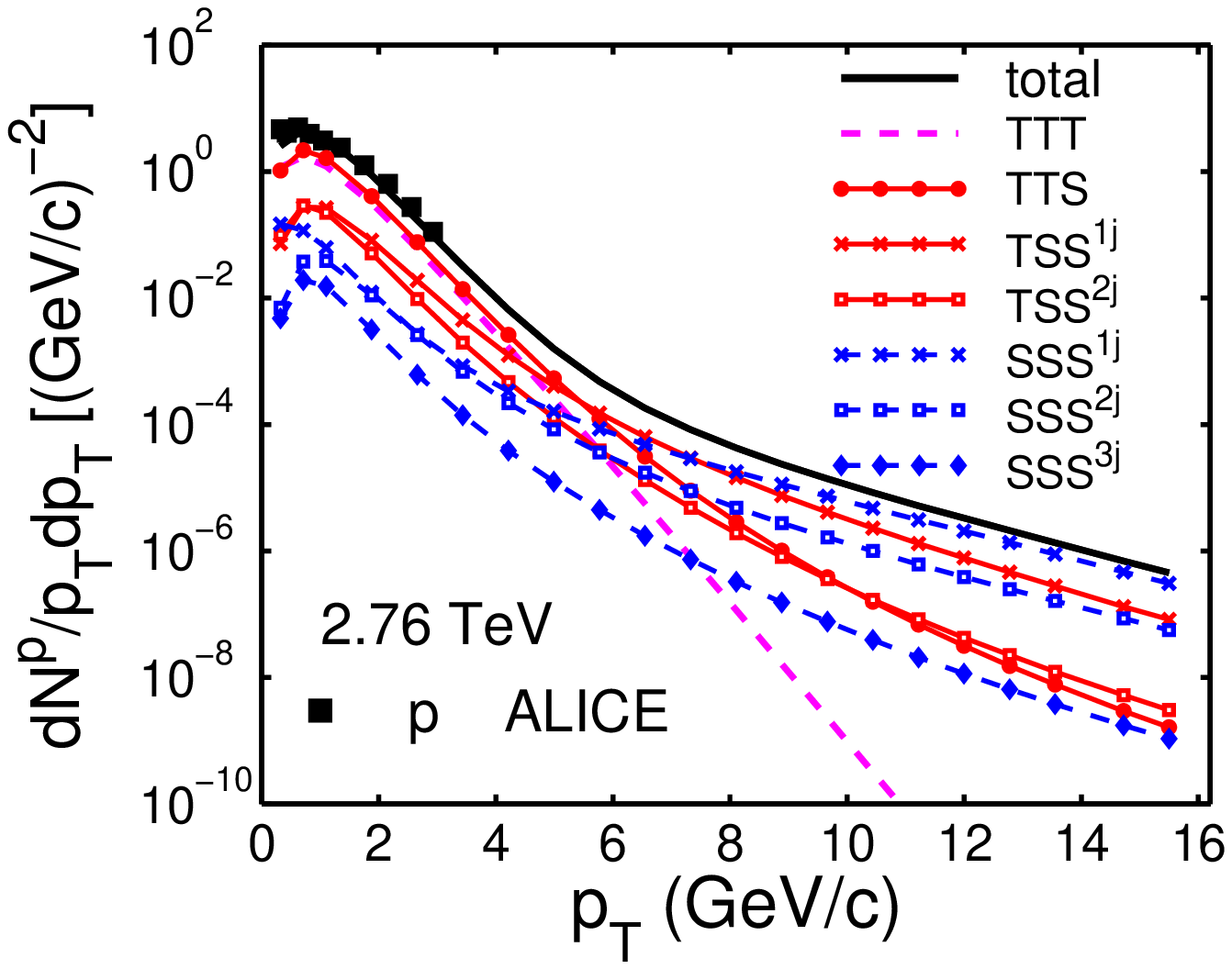}}
  \hspace{0.1in}
  \subfigure{
    \includegraphics[width=.47\textwidth]{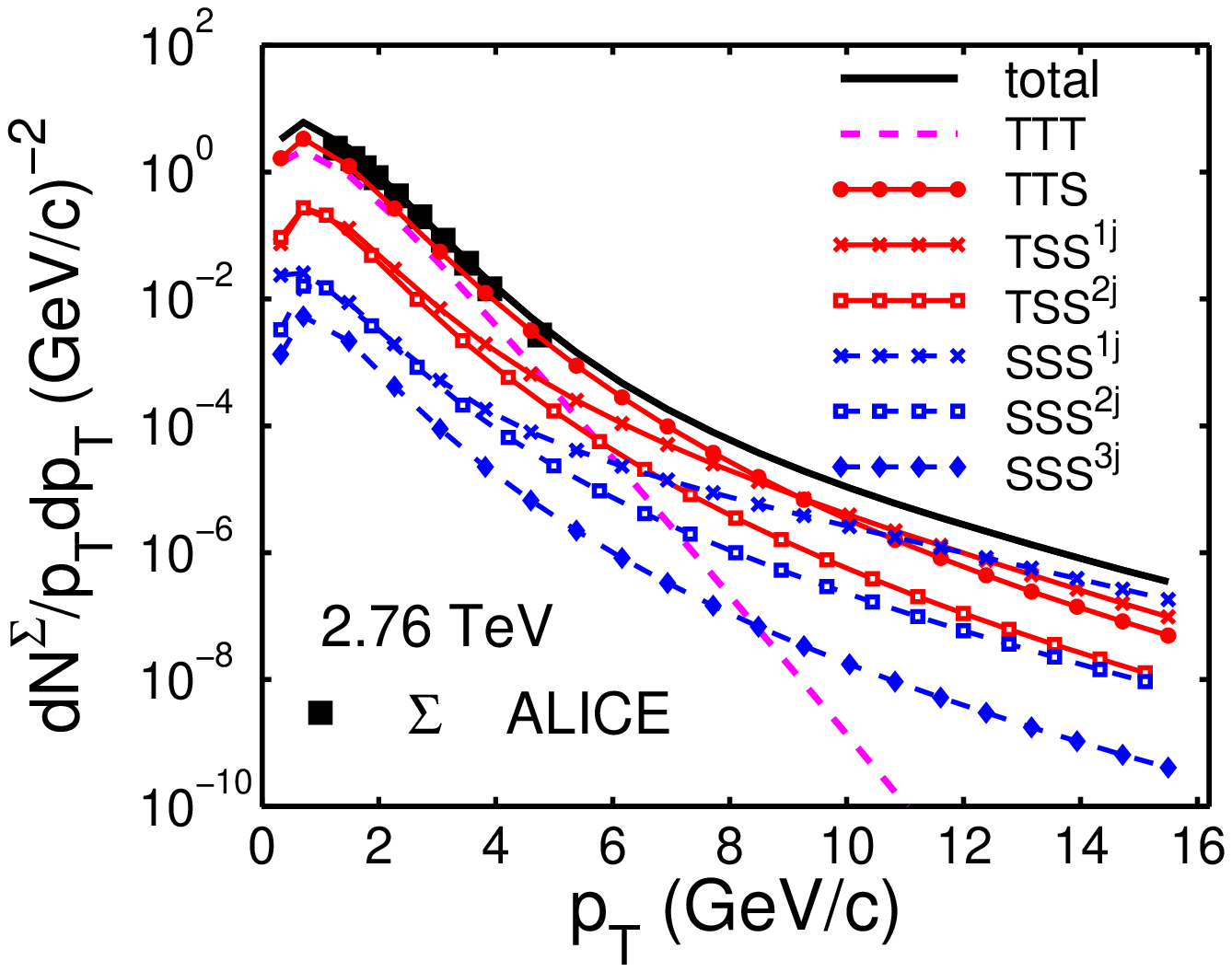}}
\caption{(Color online) Meson and baryon spectra at 2.76 TeV with various thermal and shower components shown for (a) $\pi$, (b) $K$, (c) $p$, and (d) $\Sigma$. The data are from Ref.\ \cite{mf} where $\Lambda$ distribution is shown in (d).}
\end{figure}

For mesons and baryons we show them separately in the four panels of Fig.\ 6.  The data are from Ref.\ \cite{mf} for low $p_T$, there being none at $p_T > 5$ GeV/c.  In Fig.\ 6(d) the data are for $\Lambda$ which we regard as indicative of charged hyperon $\Sigma$.  The black solid lines in each of the four panels are our predictions of the hadronic spectra for $5 < p_T < 15$ GeV/c.  One of the main points in Fig.\ 6 is the display of 2-jet contributions to the various spectra.  Note that for $\pi$ in (a) and $K$ in (b) the (SS)$^{2j}$ component is always less than (SS)$^{1j}$, but for $p$ in (c) and $\Sigma$ in (d) the (SSS)$^{2j}$ components are almost as large as (SSS)$^{1j}$ in the $2 < p_T < 6$ GeV/c region.  The ratios (SSS)$^{2j}$/(SSS)$^{1j}$ in the $p$ and $\Sigma$ spectra are shown in Fig.\ 7.  At $p_T \approx 2$ GeV/c the peaks reach as high as $\sim 1$.  Although T(SS)$^{2j}$ is not greater than T(SS)$^{1j}$, they are approximately equal in the low $p_T$ region.  (SSS)$^{3j}$ is small enough to be neglected at all $p_T$.  The conclusion is that 2-jet contributions to baryon production, though not large, can make quantitative difference in comparing theoretical results to the data.

\begin{figure}[tbph]
\centering
\includegraphics[width=0.5\textwidth]{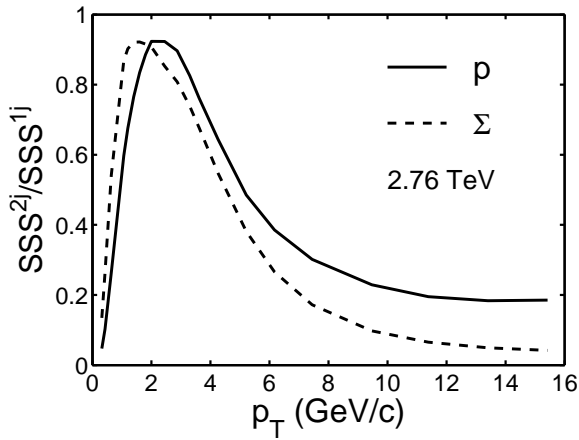}
\caption{The ratio of SSS recombination in two jets to that in one jet for $p$ (solid) and $\Sigma$ (dashed) production at 2.76 TeV.}
\end{figure}

To find a phenomenological description of minijets, we recall the experimental way discussed in Refs.\ \cite{lr,tt,tt1,tt2}, where  transverse
rapidity is introduced, defined by
\begin{eqnarray}
y_t=\ln [(m_T+p_T)/m_\pi] \ ,     \label{6.1}
\end{eqnarray}	
where $m_\pi$ is used in the denominator and in $m_T$, by definition, whether or not the hadron refers to a pion or a proton. In those references a
peak in $y_t$ is found that is attributed to minijets. It can be modeled by a Gaussian distribution centered at $y_t\approx 2.7$. Since
$dy_t/dp_T=1/m_T$, the distribuitons $dN_h/p_Tdp_T$ that we have determined can be readily transformed to $dN_h/y_tdy_t$.  Leaving out the TT and TTT components, we show them in Fig.\ 8 for $\pi$ and $p$.  We see that various components involving S indeed show peaks in $y_t$ and that the total of all those components add up to what appear as Gaussians centered at $y_t\approx 2.2$ for  pion and $\approx 2.8$ for proton.
The rise and fall of the distributions in $y_t$ are partly due to the  definition of $y_t$. The Jacobian relating $dN_h/y_tdy_t$ to $dN_h/p_Tdp_T$ is
$J=m_Tp_T/y_t$, which vanishes as $p_T\to 0$, so even a pure exponential in $dN/p_Tdp_T$ would show a bump in $dN/y_tdy_t$. The peaks we see in
Fig.\ 8 are, however, due more significantly to the suppression of low-$p_i$ shower partons. As discussed in \cite{hz}, the energy loss of
semihard partons while traversing the medium leads to the enhancement of thermal partons, so the partition between thermal and shower partons at low
$p_i$ is model dependent. The cut-off factor $\gamma_2(p_1)$ given in Eq.\ I-(A5) marks the end of the shower partons at low $p_1$, resulting in the
suppression of all TS, SS, TTS, TSS and SSS components at low $p_T$, and hence low $y_t$.

It is evident from Fig.\ 8(a) that (SS)$^{1j}$ is small compared to TS in $\pi$ production, and that (SS)$^{2j}$ is even smaller. For $p$ proton production we see from Fig.\ 8(b) that TTS is dominant, but TSS$^{2j}$ is of comparable magnitude compared to TSS$^{1j}$, though both are small compared to TTS. The components SSS$^{nj}$ are significantly smaller in the $y_t$ region shown, though dominant for $p_T>6$ GeV/c.

\begin{figure}
  \centering
  \subfigure{
    \includegraphics[width=.47\textwidth]{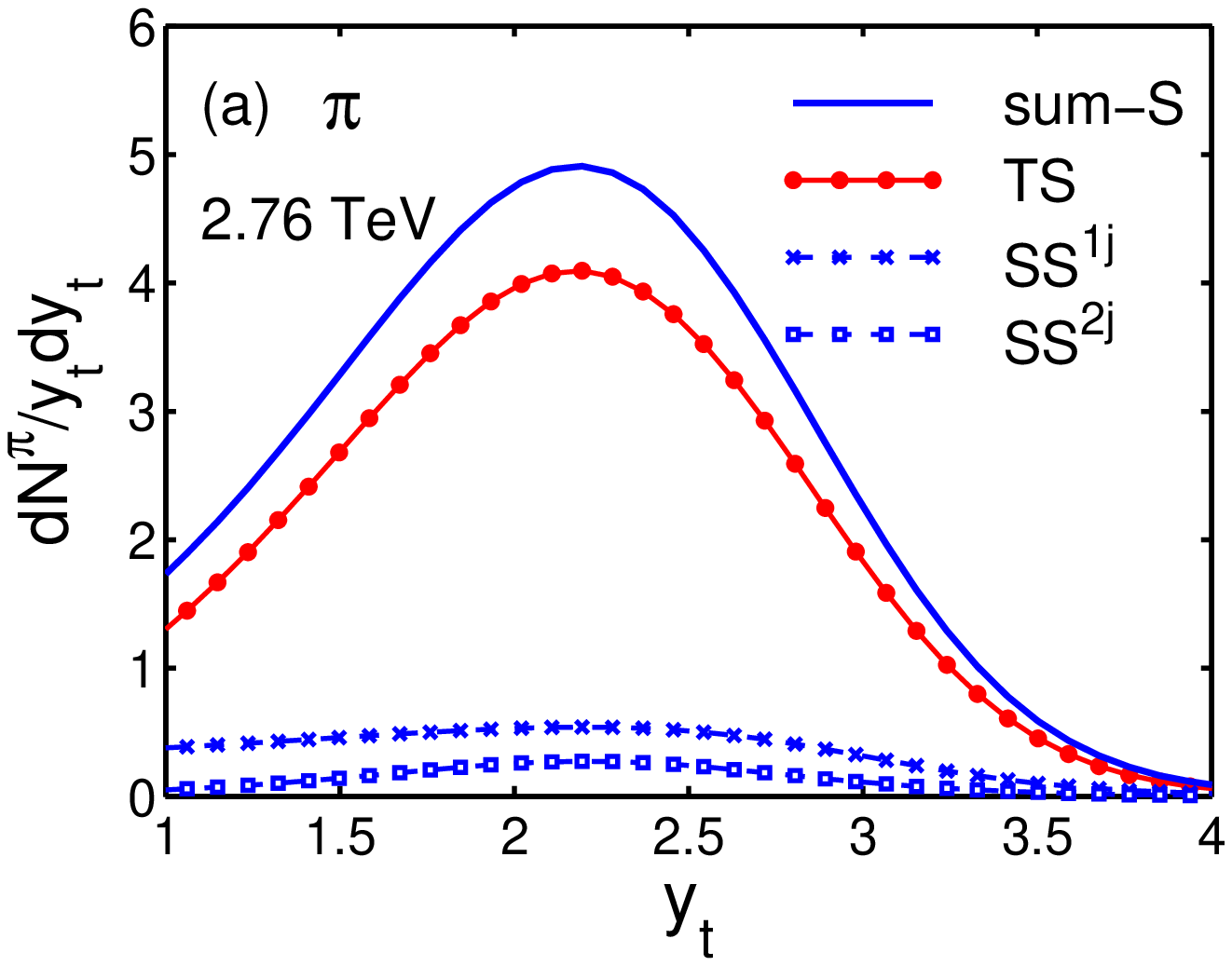}}
  \hspace{0.1in}
  \subfigure{
    \includegraphics[width=.47\textwidth]{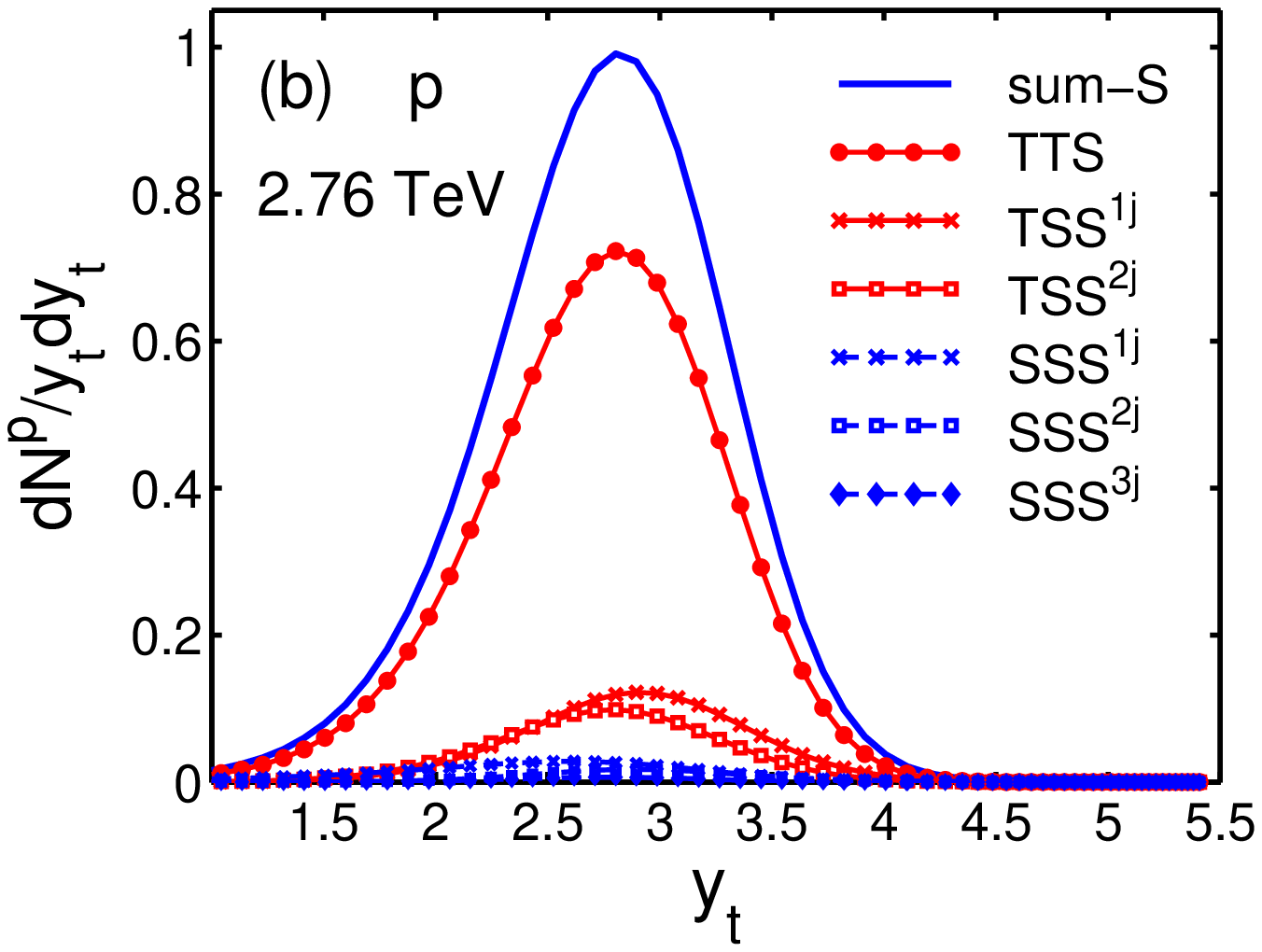}}
  \caption{(Color online) Distributions in transverse rapidity $y_t$ for $\pi$ and $p$ production at 2.76 TeV.}
\end{figure}

 \begin{figure}[tbph]
\centering
\includegraphics[width=0.5\textwidth]{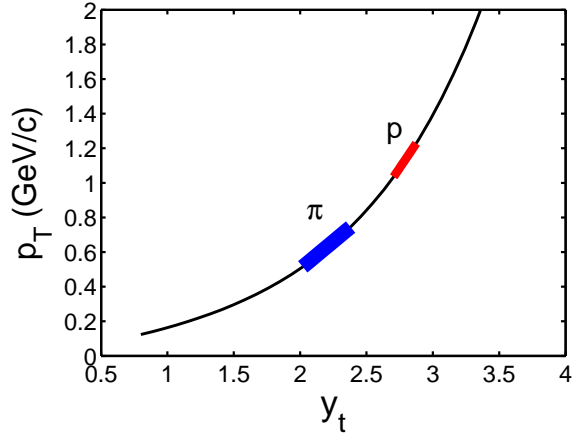}
\caption{(Color online) $p_T$ vs $y_t$ showing pion peak in wider (blue) region and proton peak in narrower (red) region.}
\end{figure}

The $y_t$ distributions in Fig.\ 8 have the virtue of exhibiting clearly the relative magnitudes of the different components in the peak region,  more so than in  $p_T$.
The relationship between $p_T$ and $y_t$ is shown in Fig.\ 9 where the shaded regions correspond to where the peaks of $\pi$ and $p$ in $y_t$ occur. It is then clear that those peaks are at $p_T\approx 0.6$ and 1.1 GeV/c, respectively,
 although they  cannot be easily identified in the plots of $dN^h/p_Tdp_T$.

It is informative to compare the magnitudes of the minijet contributions to $\pi$ and $p$ with the contributions from thermal partons only. The former are summarized by sum-S depicted by solid blue lines in Figs.\ 8(a) and (b) and are reproduced by the same line type in Fig.\ 10, in which the latter are represented by the purple dashed lines [TT for pion without resonance in (a) and TTT for proton in (b)]. The total of all components are the thick black lines in Fig.\ 10. Note that TT distribution in $y_t$ has a dip at $y_t=0$ even though its $p_T$ distribution is exponential. Except for $y_t<1.3$, sum-S is larger than TT and TTT, although for the latter the proton RF suppresses all components. The implication of this result is that the effect of minijets on hadronic spectra is dominant over essentially all $p_T$.

\begin{figure}
  \centering
  \subfigure{
    \includegraphics[width=.47\textwidth]{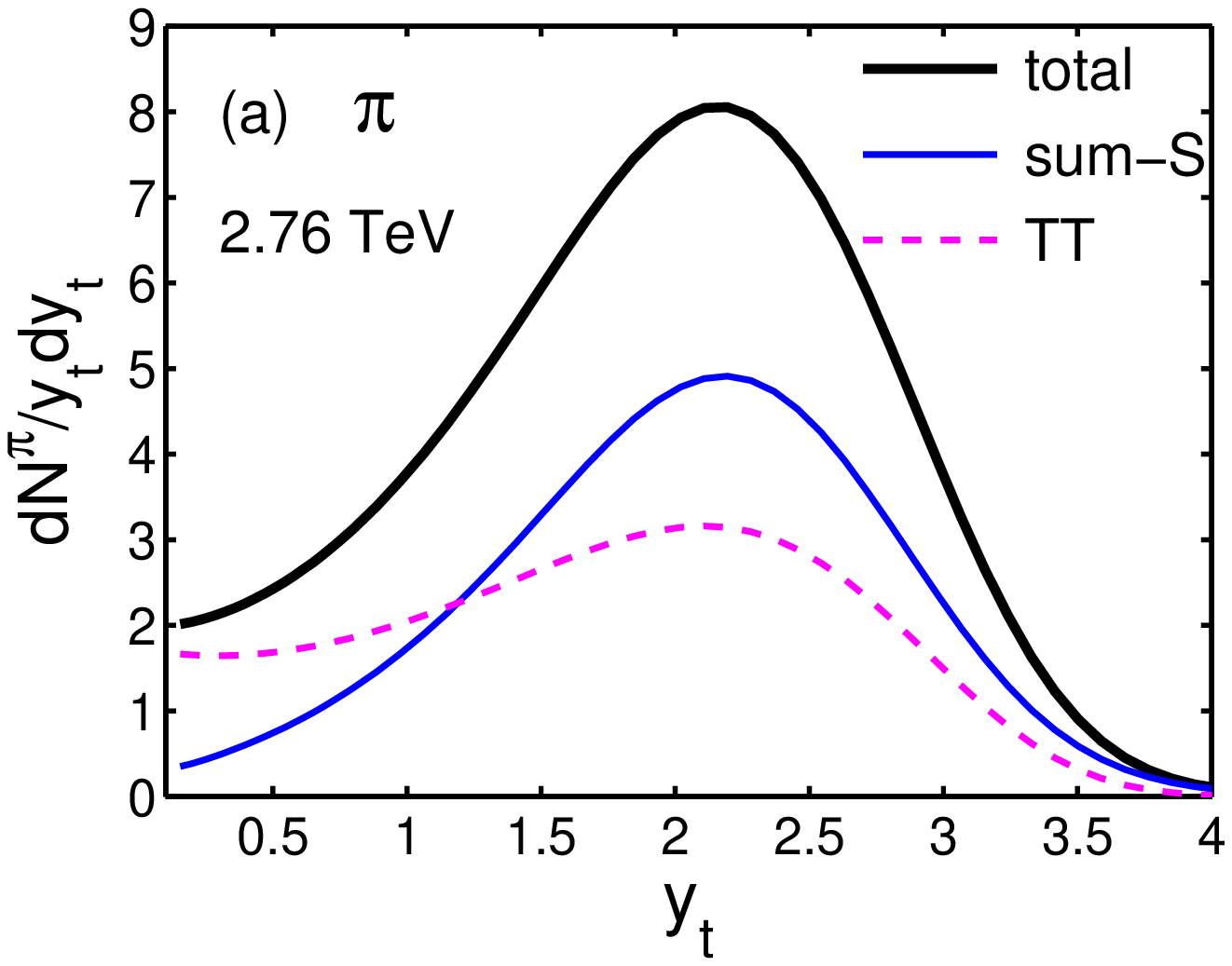}}
  \hspace{0.1in}
  \subfigure{
    \includegraphics[width=.47\textwidth]{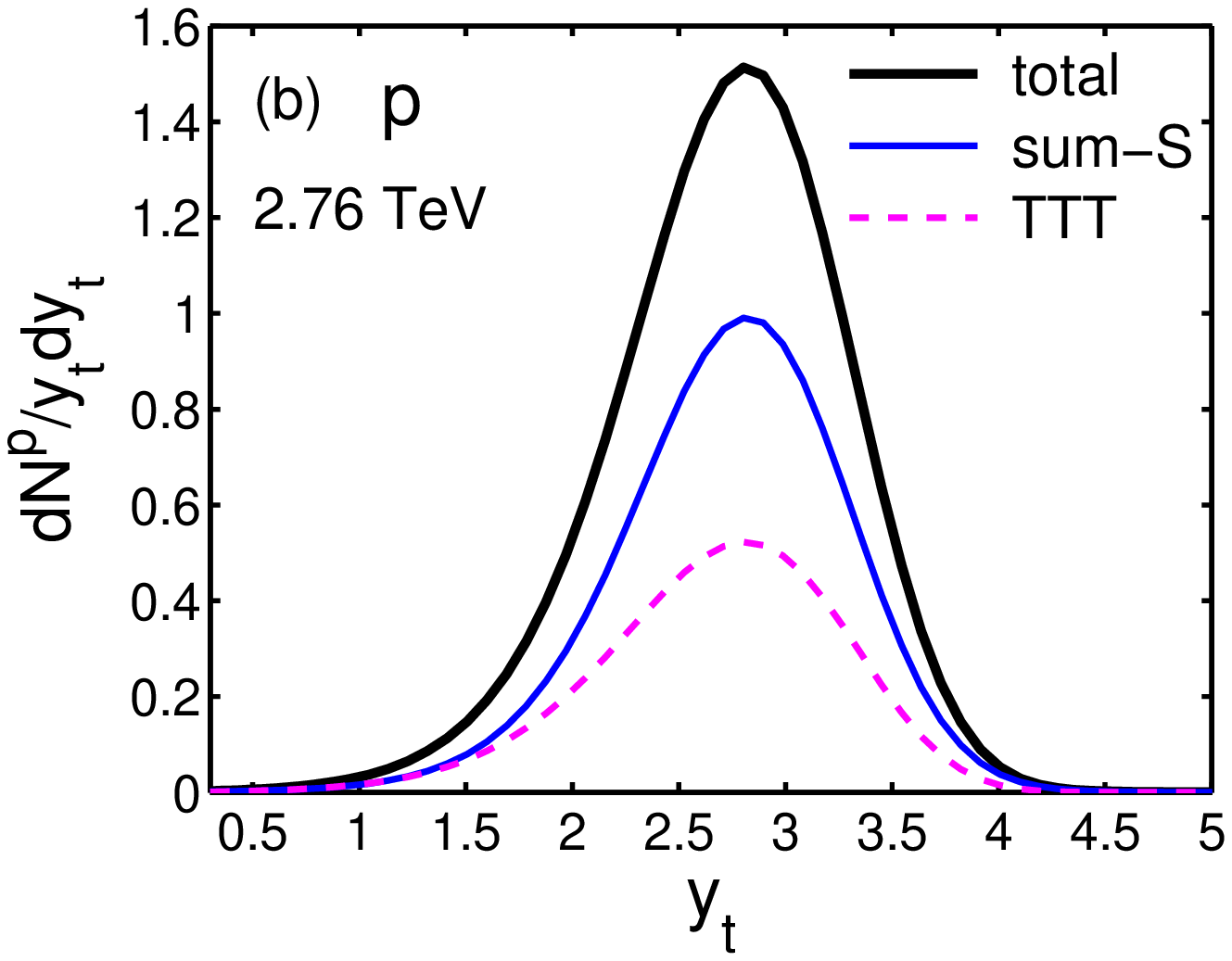}}
  \caption{(Color online) $y_t$ distributions at 2.76 TeV for (a) $\pi$ and (b) $p$, showing sum of all S components (blue solid) compared to pure thermal components (purple dashed). Thicker (black) solid lines are the totals.}
\end{figure}

In the two-component model \cite{tt,tt1} a separation of the hadronic spectra into soft and hard components is carried out by demanding that the soft
component has a specific form, viz., a L\'evy distribution in $m_T$. What remains after the subtraction of that is the hard component, which is characterized
(apart from a centrality factor) by a universal function $H_{NN}(y_t)$ that is Gaussian in $y_t$, exhibiting a peak at $y_t\sim 2.7$. The region
around the peak is regarded as being dominated by minijets. The separation is done at the hadronic level, working with observed data on pion and
proton production at RHIC. What we have done in this article is to work at the partonic level, finding many ways in which the shower partons can
contribute to hadrons, viz., in terms of TS, SS, etc. When the resultant hadronic distributions  without the soft TT components are plotted in $y_t$, we find
the peak originally found in Refs. \cite{tt,tt1}. Thus there is phenomenological agreement on what may be identified as minijet, although the avenues
of approaches to that common ground are quite different. In particular,
we assume the thermal parton distribution to be exponential that can give a good description of both pions and protons at low $p_T$ when combined with sum-S. We calculate the shower
parton distribution but use a cut-off to keep the unreliable part at very low $p_i$ from exceeding the thermal distribution.
 Our hadronization scheme treats fragmentation as a part of recombination (SS and SSS), and includes cross terms TS, TTS and TSS explicitly.
 Furthermore, by working with the transverse momenta $p_i$ of the partons we have additivity that yields the hadronic $p_T$ directly. It is a simple
 property that relates partons to hadrons, but
is lost in $y_t$.
It should finally be noted that, since the TT and TTT components in Fig.\ 10 show peaks in $y_t$ without any contributions from shower partons, any claim on the existence of minijets cannot be made convincingly without exhibiting correlation, a
subject to be considered in Sec.\ VIII.

  \section{Hadronic Spectra at 5.5 TeV}
At $\sqrt{s_{NN}} = 5.5$ TeV  the hard parton distribution $f_i(k)$ at creation is changed from what it is at lower energy and is parametrized in Ref.\
\cite{sgf}.  Other parameters, notably $\kappa_0$ and $\kappa_1$, may differ from those given in Eq.\ (\ref{5.3}), but without data they cannot be
determined.  To see the relative importance of the different components, it is useful to calculate them for 5.5 TeV without assuming any specific
changes of those parameters.  The results on the $\pi$ and $p$ spectra
are shown in Figs.\ 11(a) and (b).  In comparing Fig.\ 11(a) to Fig.\ 6(a) for pion production we note that while the
behavior at low $p_T$ is unchanged due to the assumption of unaltered thermal partons, the distribution at the high-$p_T$ end increases by roughly an
order of magnitude.  Furthermore,  ({SS})$^{2j}$ become almost as large as TS for $p_T\ ^>_\sim\ 6$ GeV/c.
\begin{figure}
  \centering
  \subfigure{
    \includegraphics[width=.47\textwidth]{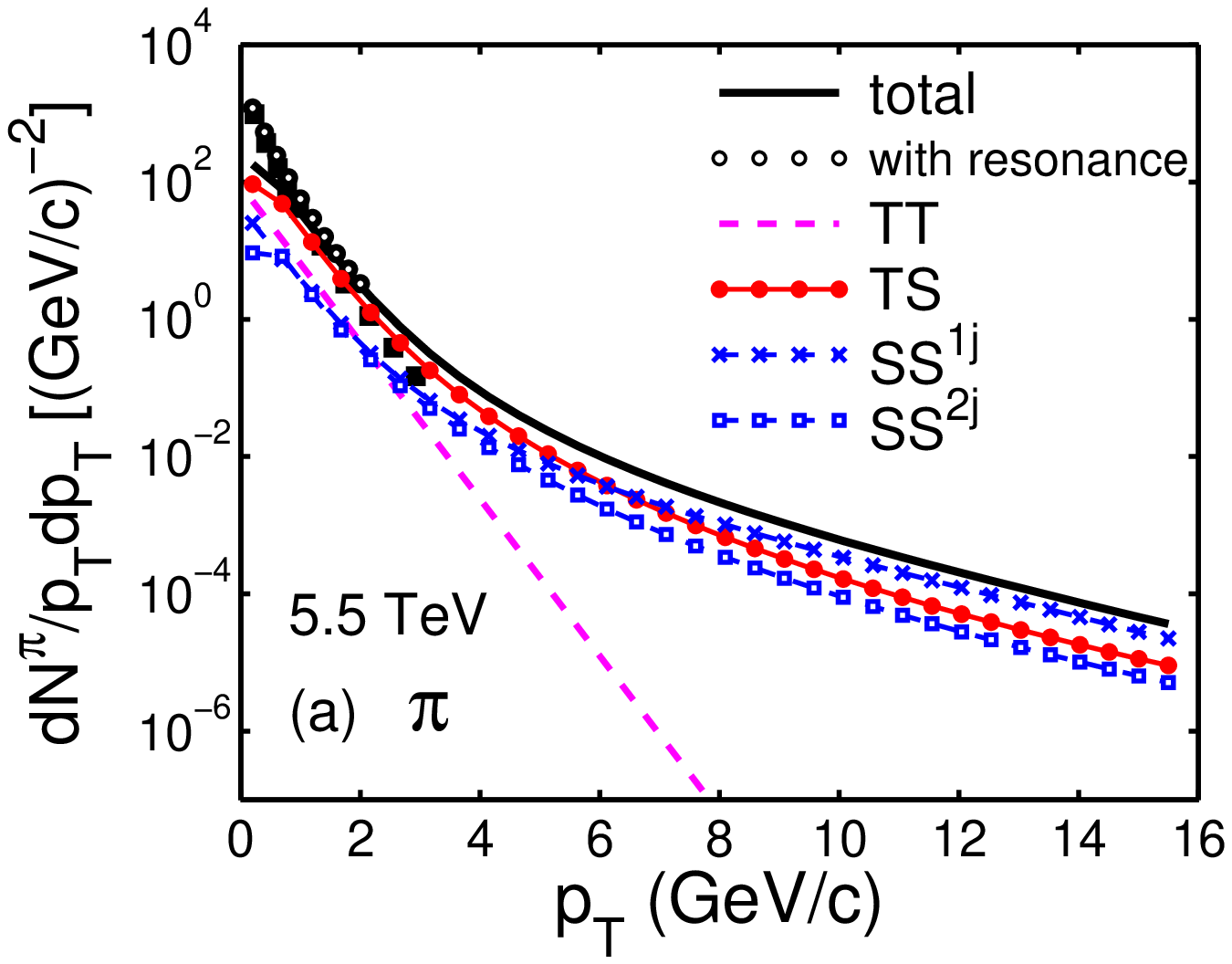}}
  \hspace{0.1in}
  \subfigure{
    \includegraphics[width=.47\textwidth]{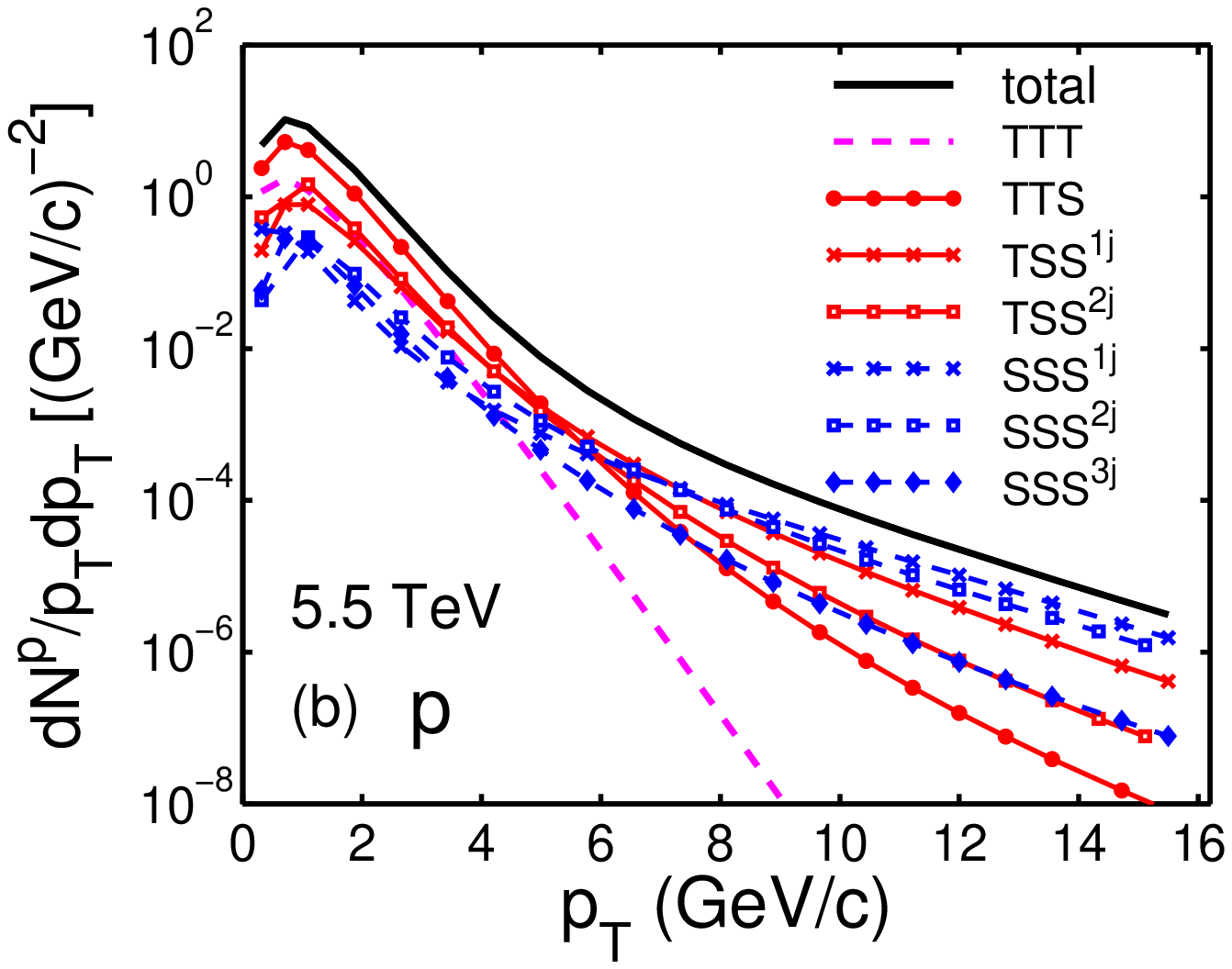}}
  \caption{(Color online) $p_T$ distributions for (a) $\pi$ and (b) $p$ at 5.5 TeV.}
\end{figure}

 A more visibly apparent change is seen in proton production by comparing Fig.\ 11(b) to Fig.\ 6(c).  The ({SSS})$^{3j}$ component increases by almost two
 orders of magnitude at $p_T \sim 16$ GeV/c.  Even at $p_T \sim 2$ GeV/c its value at 5.5 TeV is enhanced over that at 2.76 TeV by more than a factor of
 10.  One sees in Fig.\ 11(b) that multijet recombination is important at all $p_T$.
 To compare the various multi-minijet contributions, we show in Fig.\ 12(a) the ratio (SS)$^{2j}$/(SS)$^{1j}$ for pion production and in Fig.\ 12(b)
(SSS)$^{2j}$/(SSS)$^{1j}$ and (SSS)$^{3j}$/(SSS)$^{1j}$ for proton production. Note that the  proton peak for (SSS)$^{2j}$/(SSS)$^{1j}$ at 5.5 TeV in Fig.\ 12(b)  is more than twice higher compared to that at 2.76 TeV in Fig.\ 7.

\begin{figure}
  \centering
  \subfigure{
    \includegraphics[width=.47\textwidth]{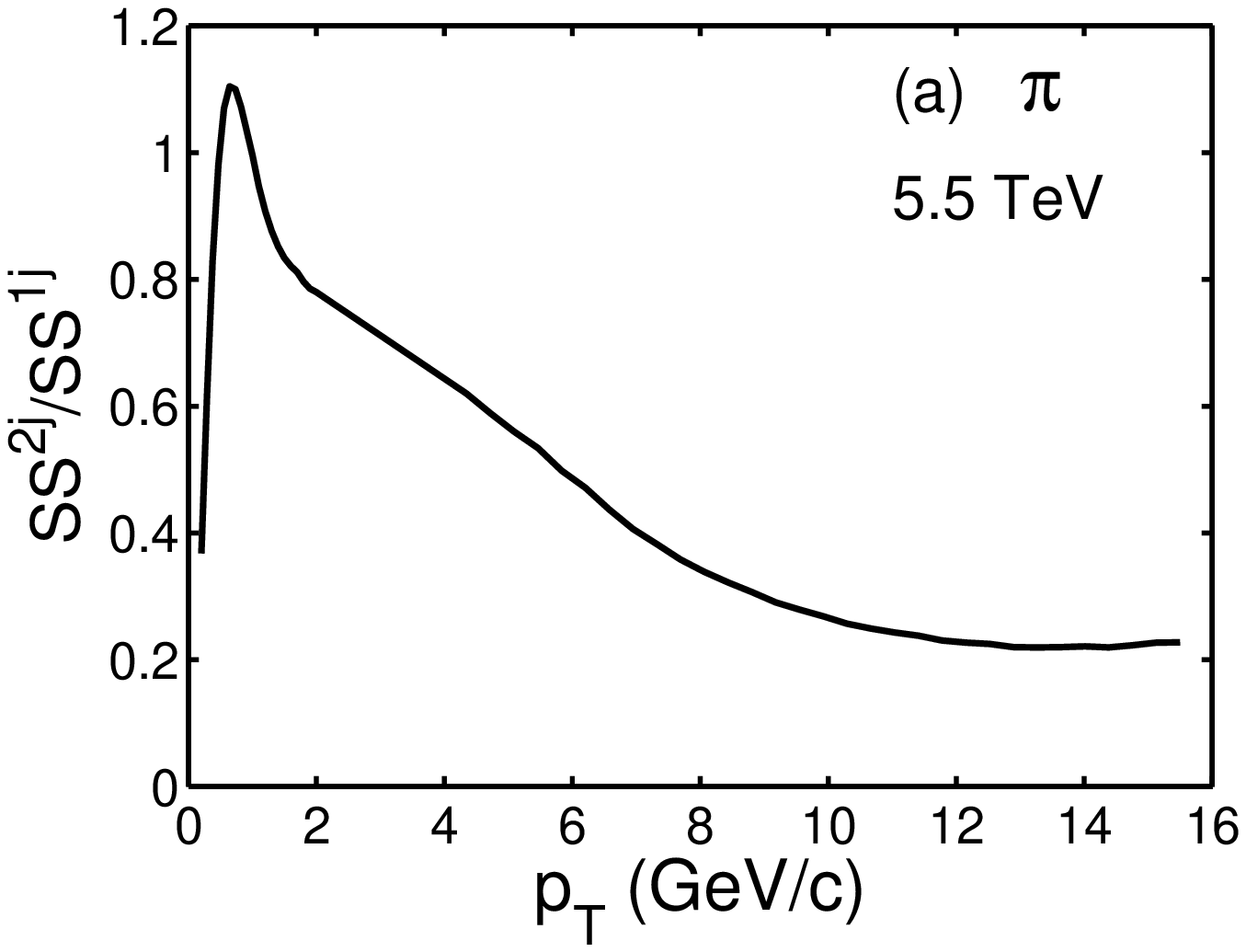}}
  \hspace{0.1in}
  \subfigure{
    \includegraphics[width=.47\textwidth]{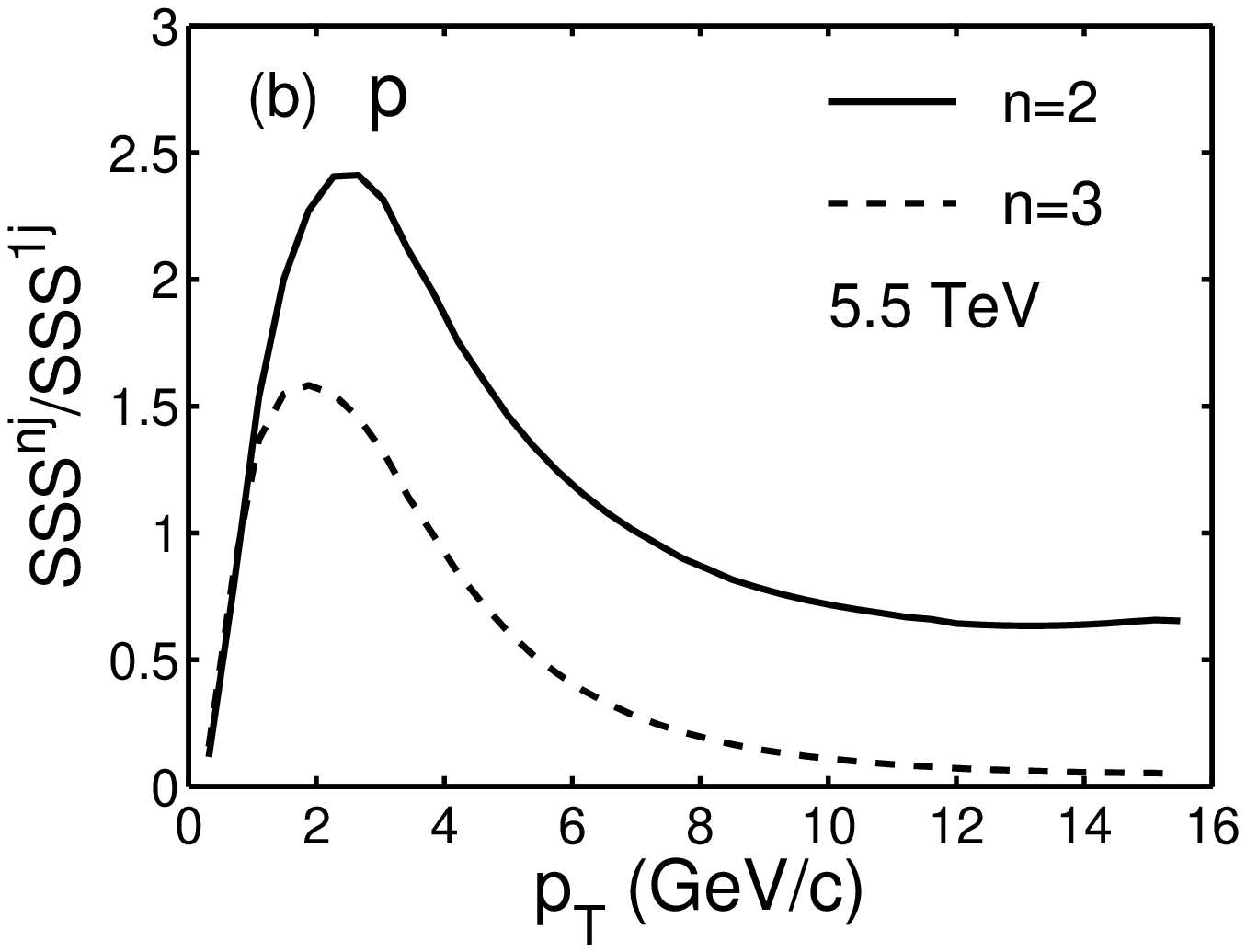}}
  \caption{The ratios at 5.5 TeV: (a) (SS)$^{2j}$/(SS)$^{1j}$ for pion production and  (b) (SSS)$^{nj}$/(SSS)$^{1j}$  for proton production with $n=2$ (solid) and $n=3$ (dashed).}
\end{figure}

\begin{figure}[tbph]
\includegraphics[width=.52\textwidth]{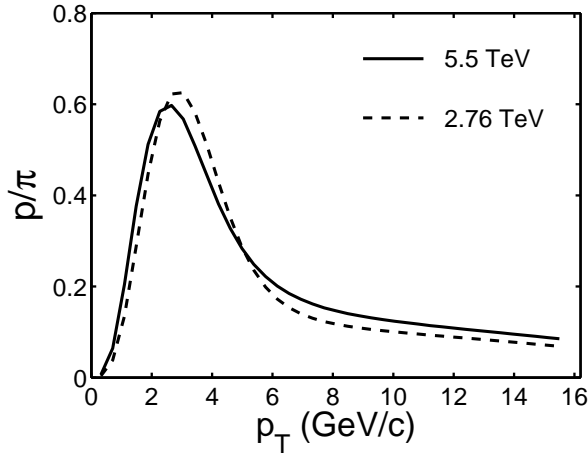}
\caption{$p/\pi$ ratio at 2.76 and 5.5 TeV.}
\end{figure}

Since both pion and proton distributions at large $p_T$ are higher at 5.5 TeV compared to those at 2.76 TeV, the $p/\pi$ ratio is not significantly
changed;  both are shown in Fig.\ 13. The peak is shifted slightly lower to around $p_T\simeq 2$ GeV/c.

\begin{figure}
  \centering
  \subfigure{
    \includegraphics[width=.47\textwidth]{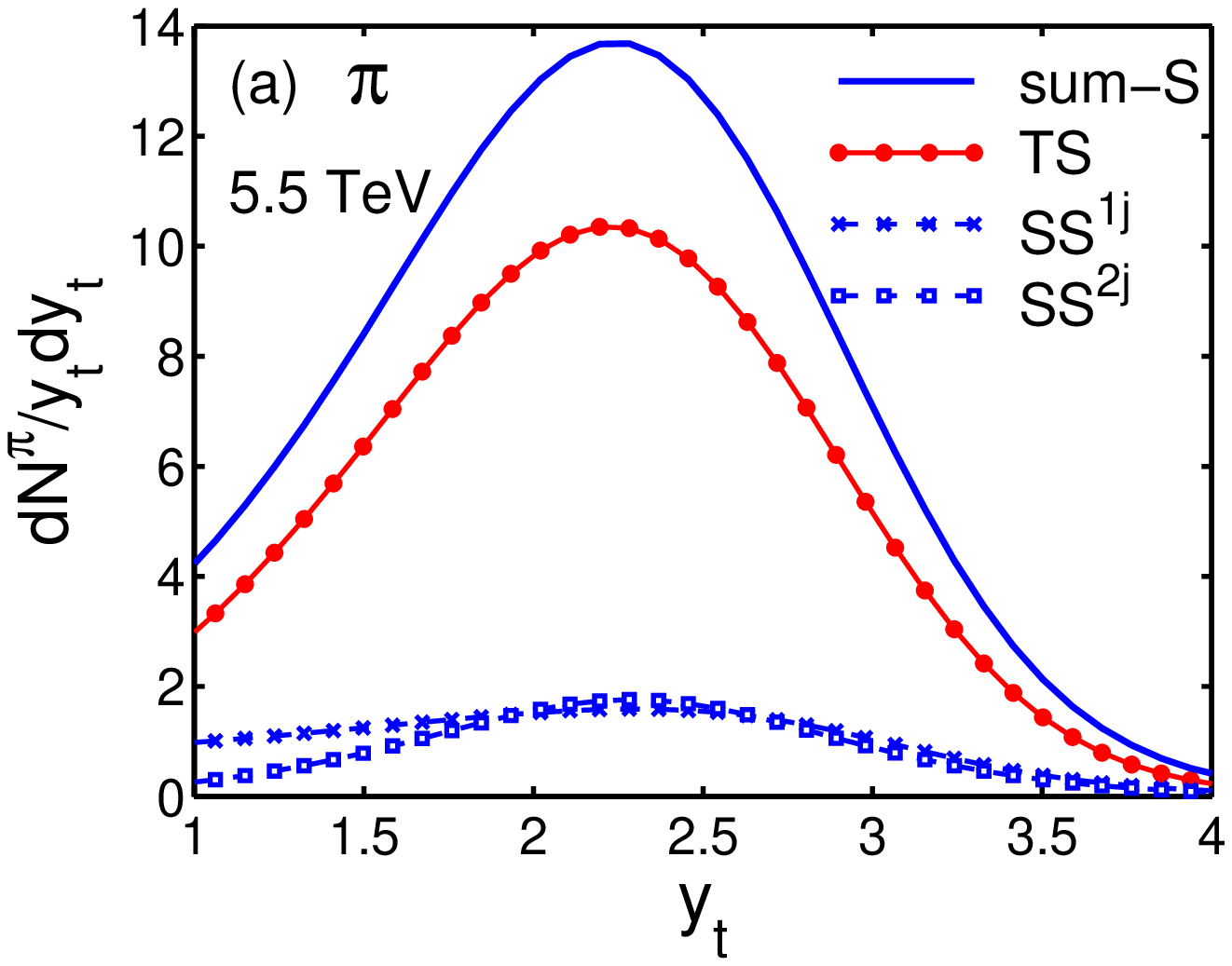}}
  \hspace{0.1in}
  \subfigure{
    \includegraphics[width=.47\textwidth]{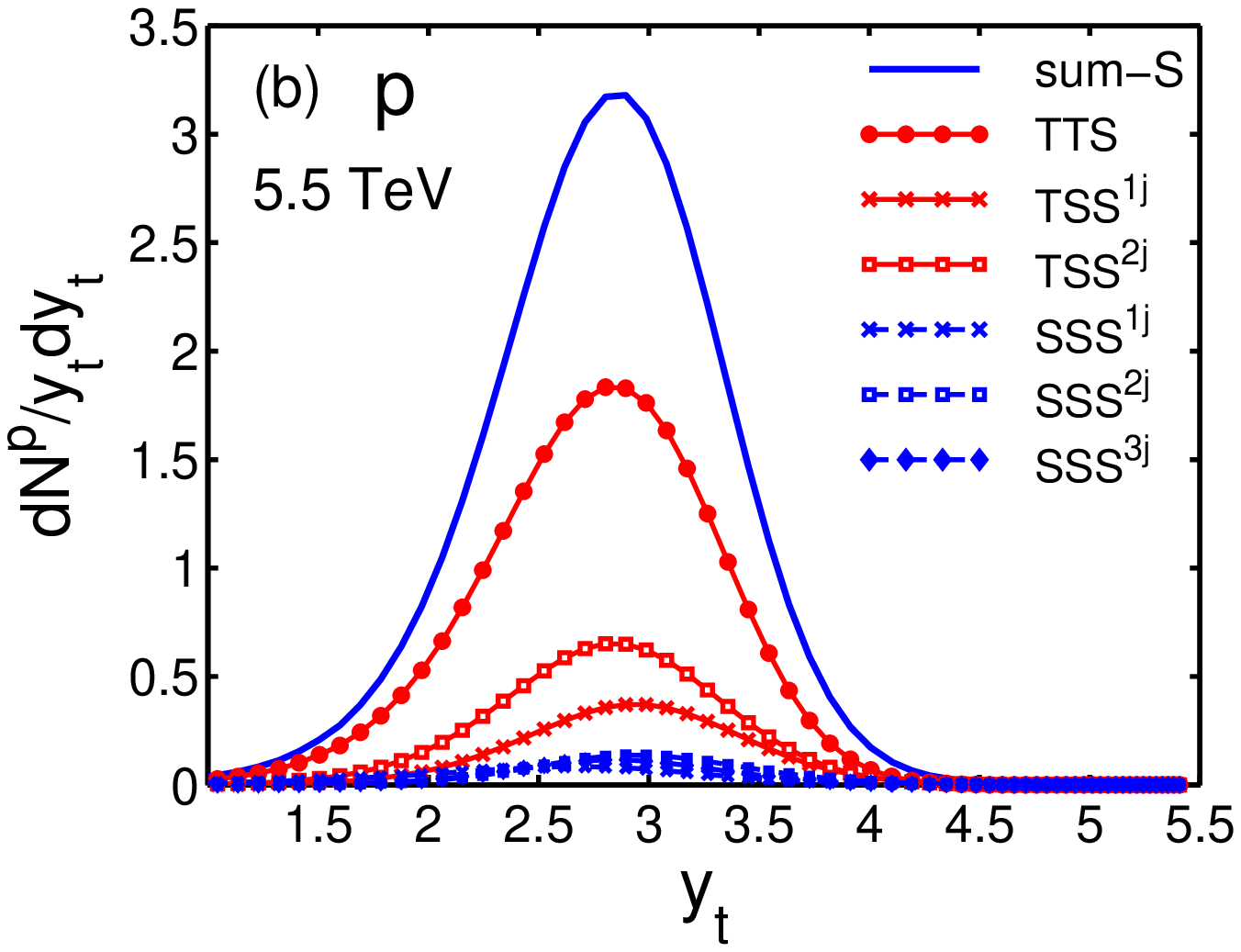}}
\caption{(Color online) Shower parton contributions to (a) pion and (b) proton \dis s in $y_t$ at 5.5 TeV.}
\end{figure}

The distributions in $y_t$ at 5.5 TeV have the general structure as at 2.76 TeV, as shown in Figs.\ 14(a) and (b) for pion and proton, respectively. The
peaks occur at around the same values of $y_t$; however, the magnitudes are much higher --- nearly triple.   Notice that (SS)$^{2j}$ is now as high as (SS)$^{1j}$, while (TSS)$^{2j}$ exceeds (TSS)$^{1j}$. At both 2.76 and 5.5 TeV, the sums of all shower parton contributions (shown by the solid blue lines in Figs.\ 8 and 14) are significantly higher than the (SS)$^{nj}$ and (SSS)$^{nj}$ contributions that do not involve the thermal partons. The differences between sum-S and pure T components are shown in Fig.\ 15(a) for $\pi$ and (b) for $p$; evidently, at 5.5 TeV sum-S is now more than 4 times larger than TT for $\pi$ at the peak, and as much as 6 times TTT for $p$. Clearly, the minijet contributions dominate over pure thermal on the one hand and over pure fragmentation on the other.

\begin{figure}
  \centering
  \subfigure{
    \includegraphics[width=.47\textwidth]{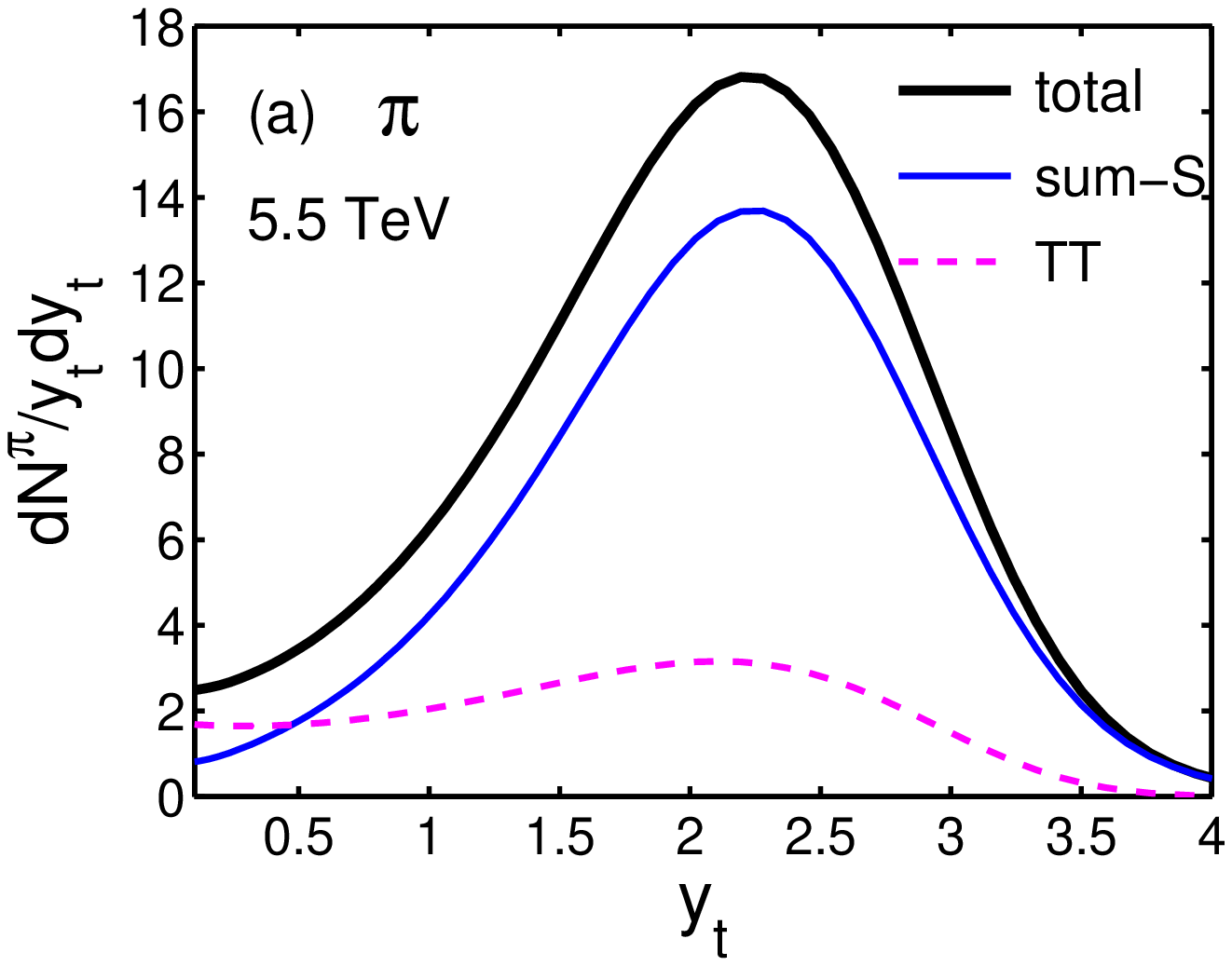}}
  \hspace{0.1in}
  \subfigure{
    \includegraphics[width=.47\textwidth]{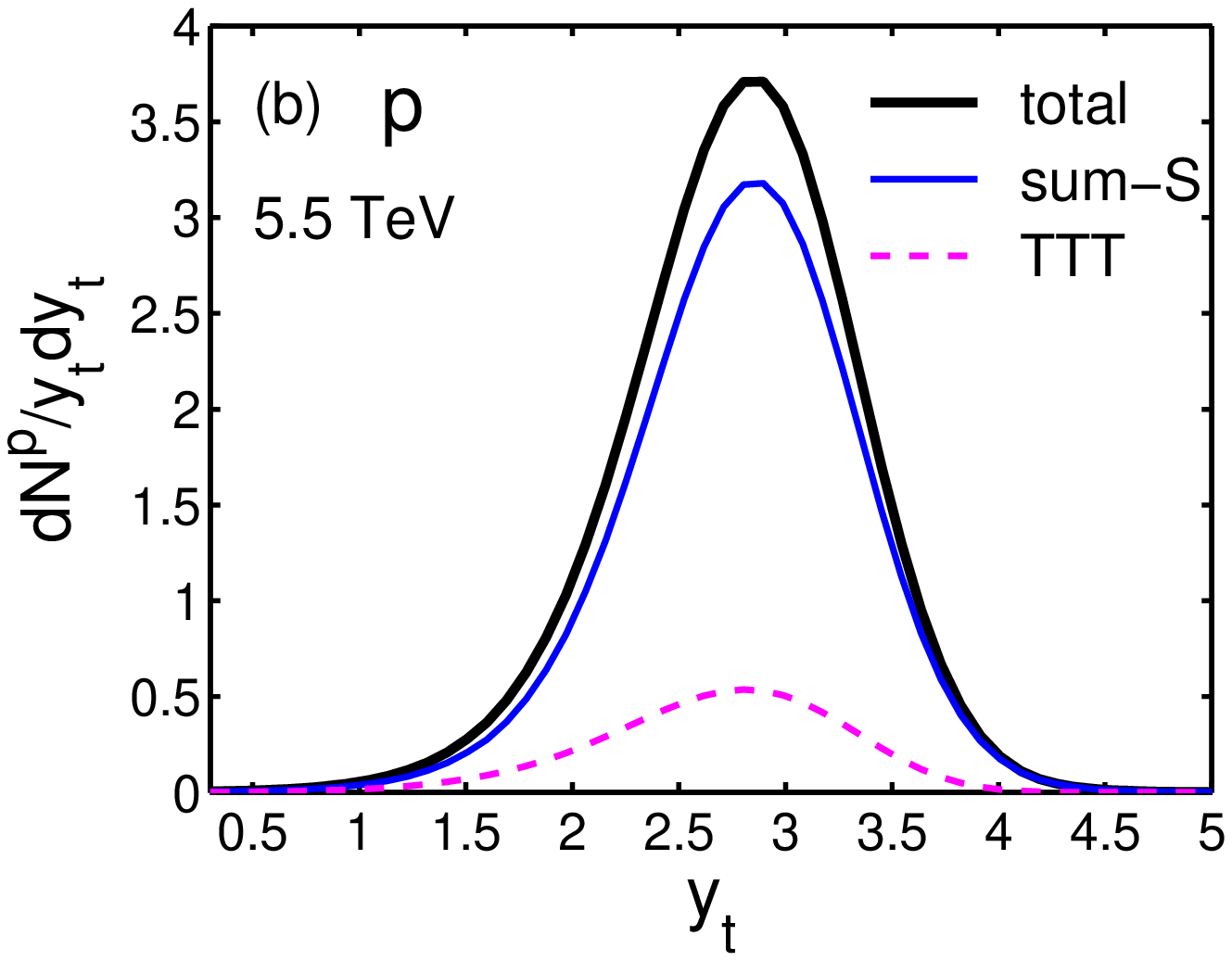}}
\caption{(Color online) Sums of S contributions compared to (a) TT and (b) TTT \dis s in $y_t$ at 5.5 TeV.}
\end{figure}

\section{Results on Two-particle Correlations}

For $\pi$-$\pi$ correlation we use Eqs.\ (A1)-(A6) to calculate $dN_{\pi\pi}/p_tp_adp_tdp_a$. There is no TT term in either the trigger or the associated particle because it is factorizable. All six terms included are shown in Fig.\ 3 and are clearly non-factorizable. The result on the correlation function $P_2^{\pi\pi}(p_t,p_a)$ defined in Eq.\ (\ref{4.1}) is shown in Fig.\ 16(a) for collisions at 2.76 TeV, exhibiting clearly that the correlation is important only when $p_t$ and $p_a$ are in the 1 GeV/c region. The most important contribution to that correlation in the peak region comes from the (TS)(TS) term in Fig.\ 3(a) that involves two shower partons produced by a single semihard parton. That cluster of partons with low transverse momenta is what has been referred to as minijet, and the correlation is among the components of that cluster, but for correlation between pions there is a gain in momentum due to recombination of the correlated shower partons with thermal partons in the vicinity. The properties near the correlation peak can best be shown in a plot of $P_2^{\pi\pi}(y_{t_1},y_{t_2})$ in Fig.\ 16(b). The peak is located at $y_{t_1}=y_{t_2}\approx 2.2$. This result is consistent with the findings at RHIC in Au-Au collisions at 200 GeV for unlike-sign correlation on the same side \cite{lr}. To show the suppressed contributions from the other components compared to (TS)(TS),
Fig.\ 17(a) exhibits (SS)$^{1j}$(SS)$^{1j}$ and (TS)(SS)$^{1j}$ relative to (TS)(TS). They are obviously insignificant for the values of $p_t$ and $p_a$ shown. For (SS)$^{2j}$ contribution to $P_2(y_{t_1},y_{t_2})$ shown in Fig.\ 17(b), they are also very small compared to that shown in Fig.\ 16(b), which is dominated by (TS)(TS).

\begin{figure}
  \centering
  \subfigure{
    \includegraphics[width=.47\textwidth]{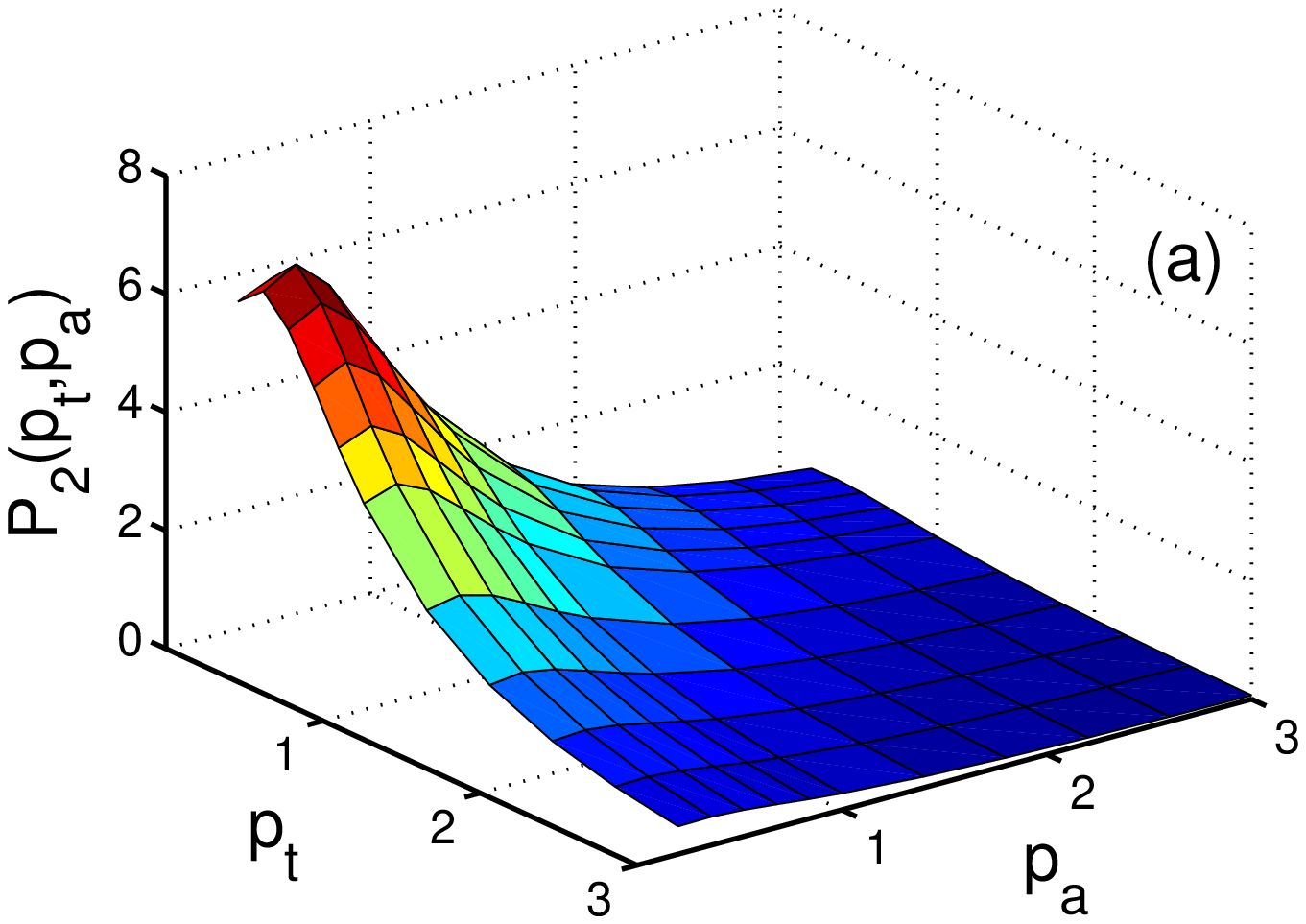}}
  \hspace{0.1in}
  \subfigure{
    \includegraphics[width=.47\textwidth]{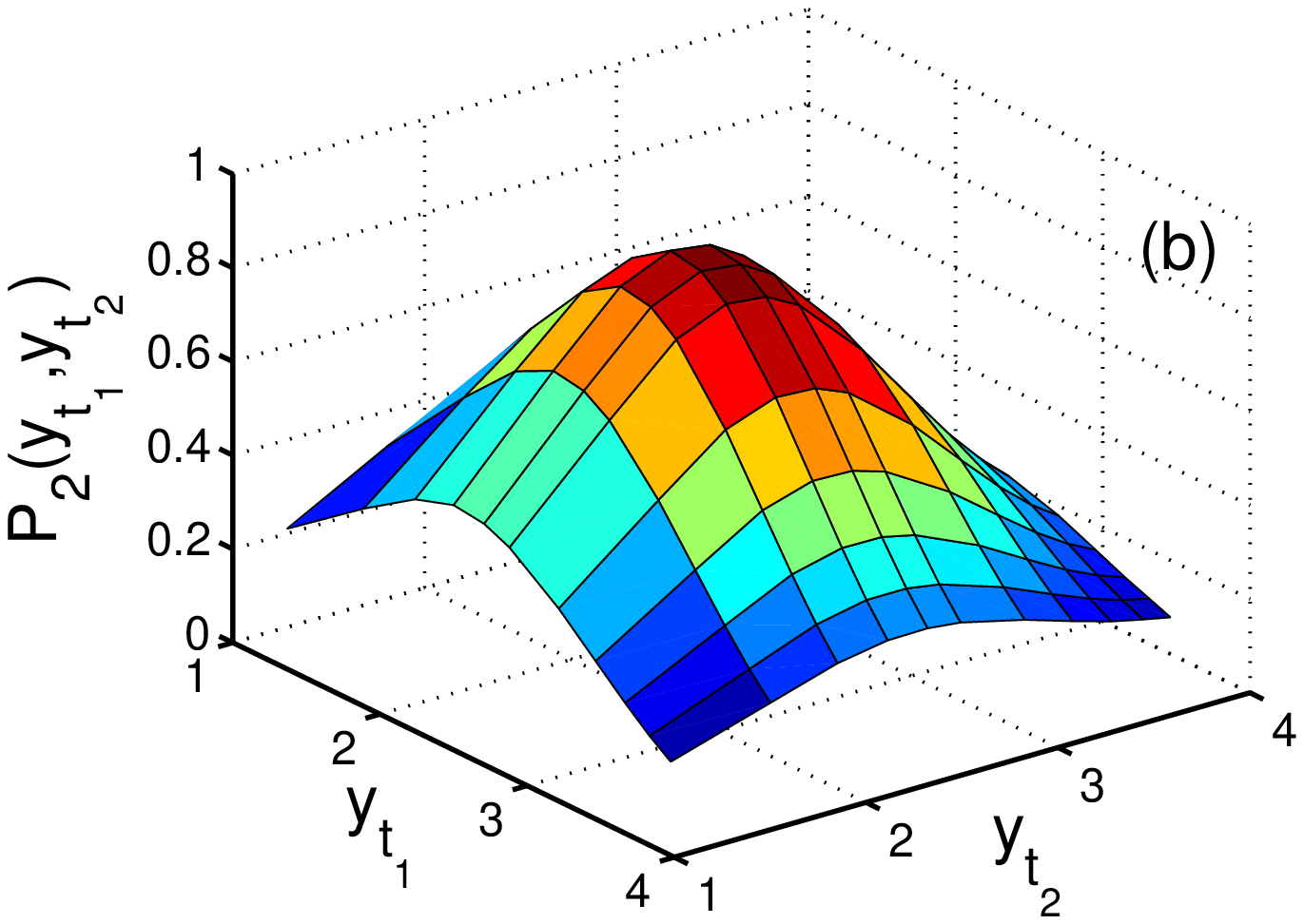}}
  \caption{(Color online) $\pi$-$\pi$ correlation at 2.76 TeV on (a) $p_t$-$p_a$ and (b) $y_{t_1}$-$y_{t_2}$}
\end{figure}

\begin{figure}
  \centering
  \subfigure{
    \includegraphics[width=.47\textwidth]{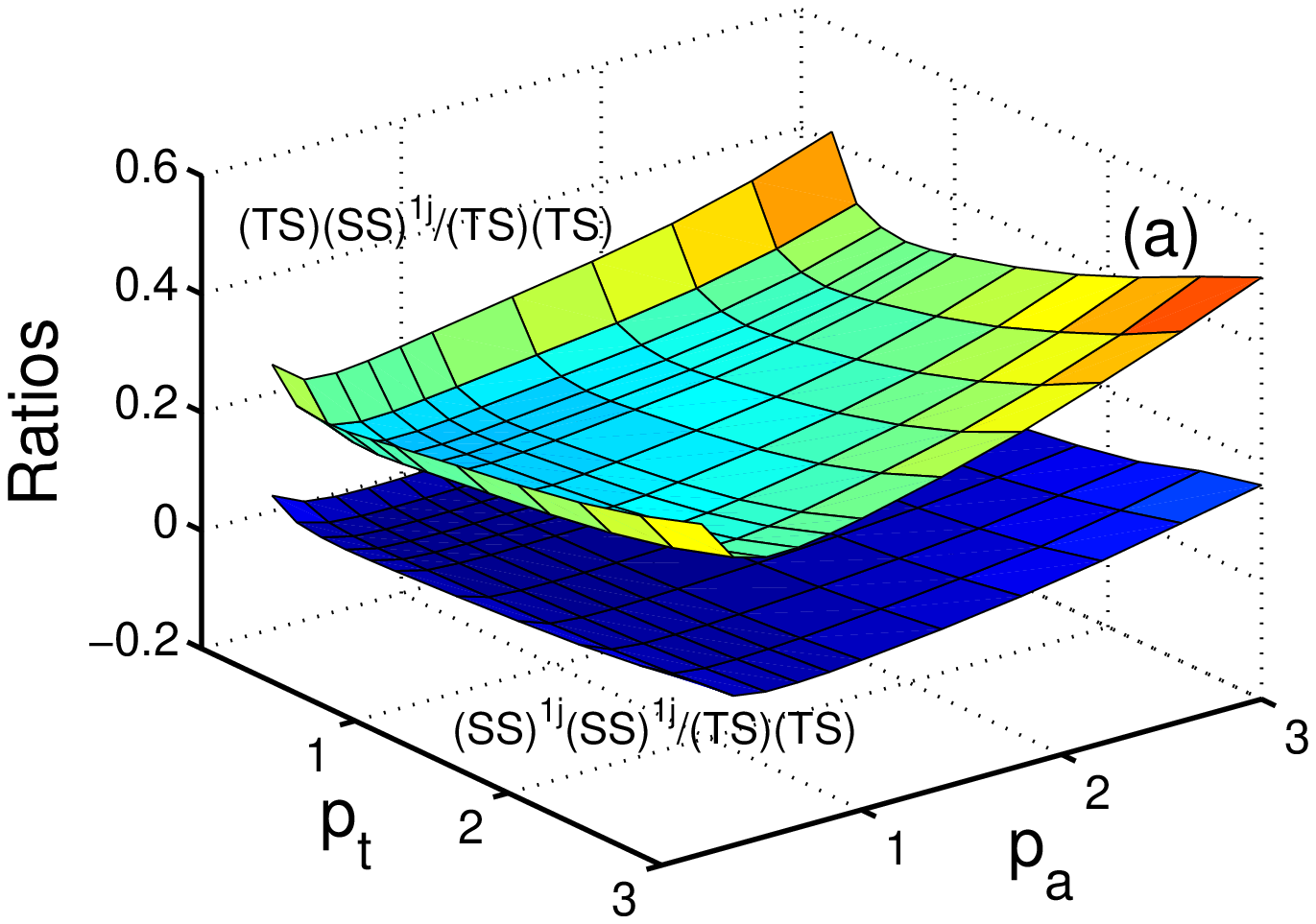}}
  \hspace{0.1in}
  \subfigure{
    \includegraphics[width=.47\textwidth]{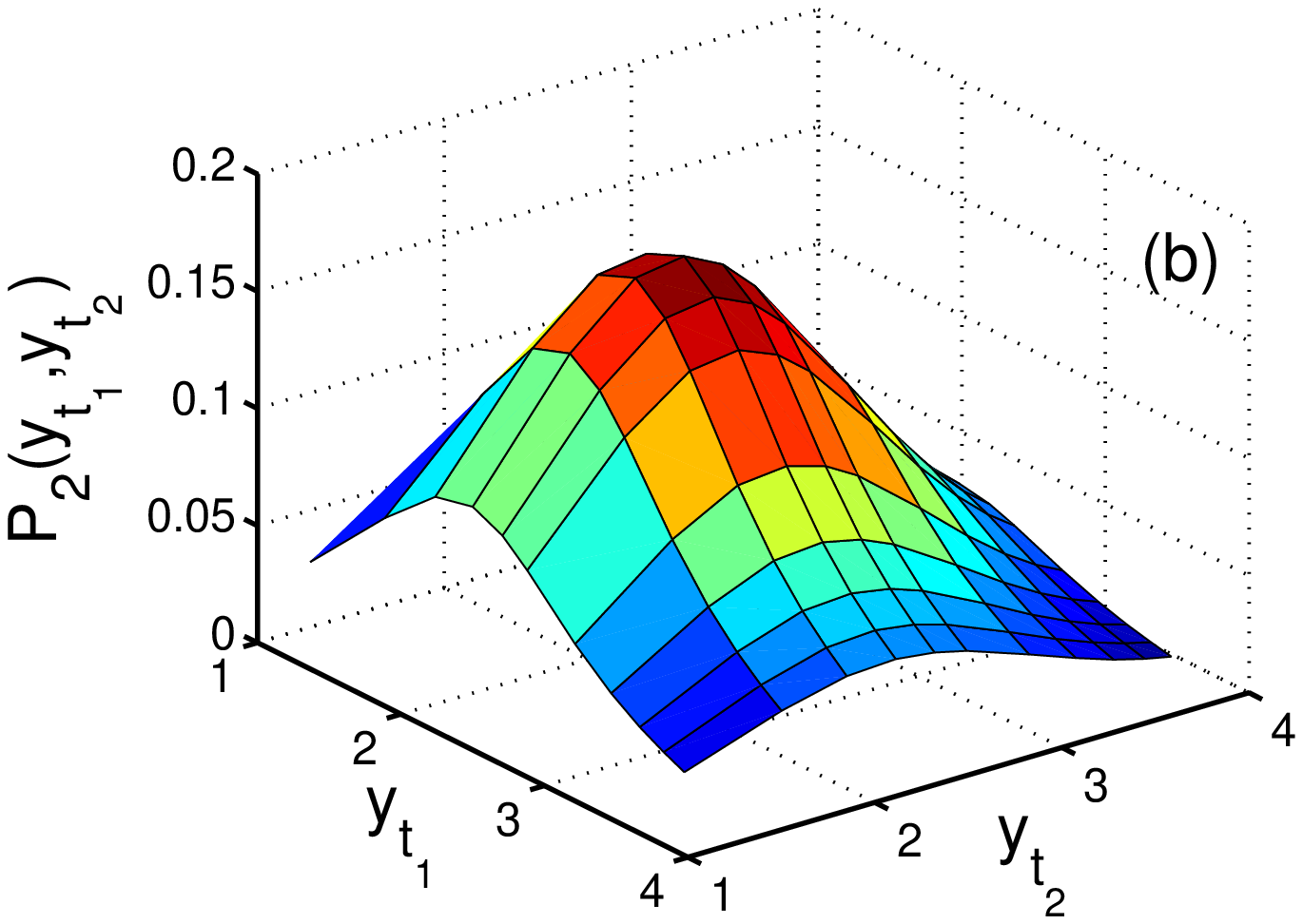}}
 \caption{(Color online) (a) Ratios of components in $\pi$-$\pi$ correlation: (TS)(SS)$^{1j}$/(TS)(TS) and (SS)$^{1j}$(SS)$^{1j}$/(TS)(TS) on $p_t$-$p_a$, (b)  $P_2(y_{t_1},y_{t_2})$ for a pion from (SS)$^{2j}$ correlated with another pion in terms shown in Fig.\ 3 (c), (e) and (f).}
\end{figure}

The correlation between two protons has even more components. We show in Fig.\ 18 only the dominant component (TTS)(TTS), as depicted in Fig.\ 4; here the same peak is shown in plots on (a) $p_t$-$p_a$ and  (b) $y_{t_1}$-$y_{t_2}$. Compared to $\pi\pi$ correlation in Fig.\ 16, the magnitude of $pp$ correlation is an order of magnitude smaller. The peak in $y_{t_1}$-$y_{t_2}$ is shifted to slightly higher value at $\approx 2.8$.

\begin{figure}
  \centering
  \subfigure{
    \includegraphics[width=.47\textwidth]{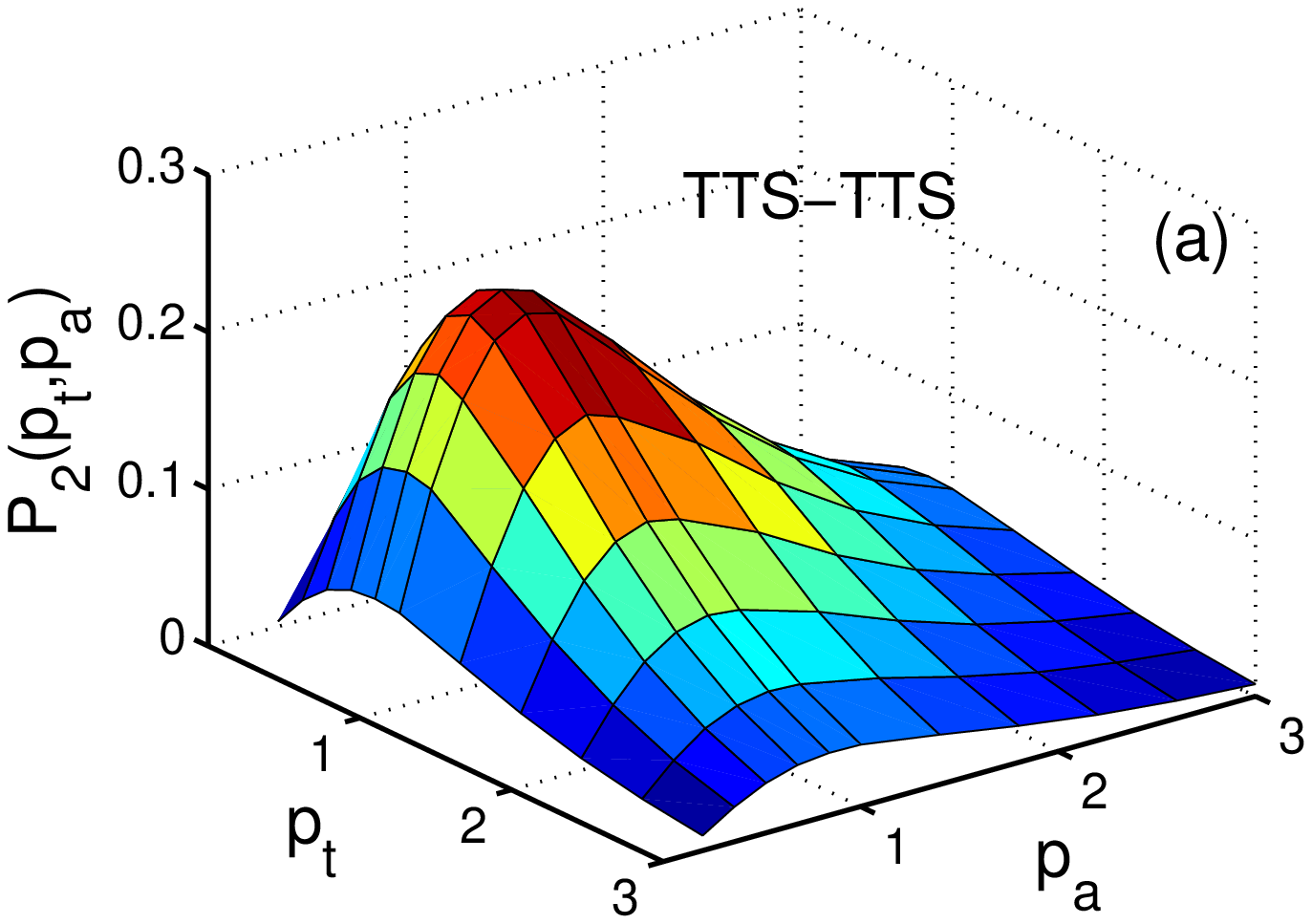}}
  \hspace{0.1in}
  \subfigure{
    \includegraphics[width=.47\textwidth]{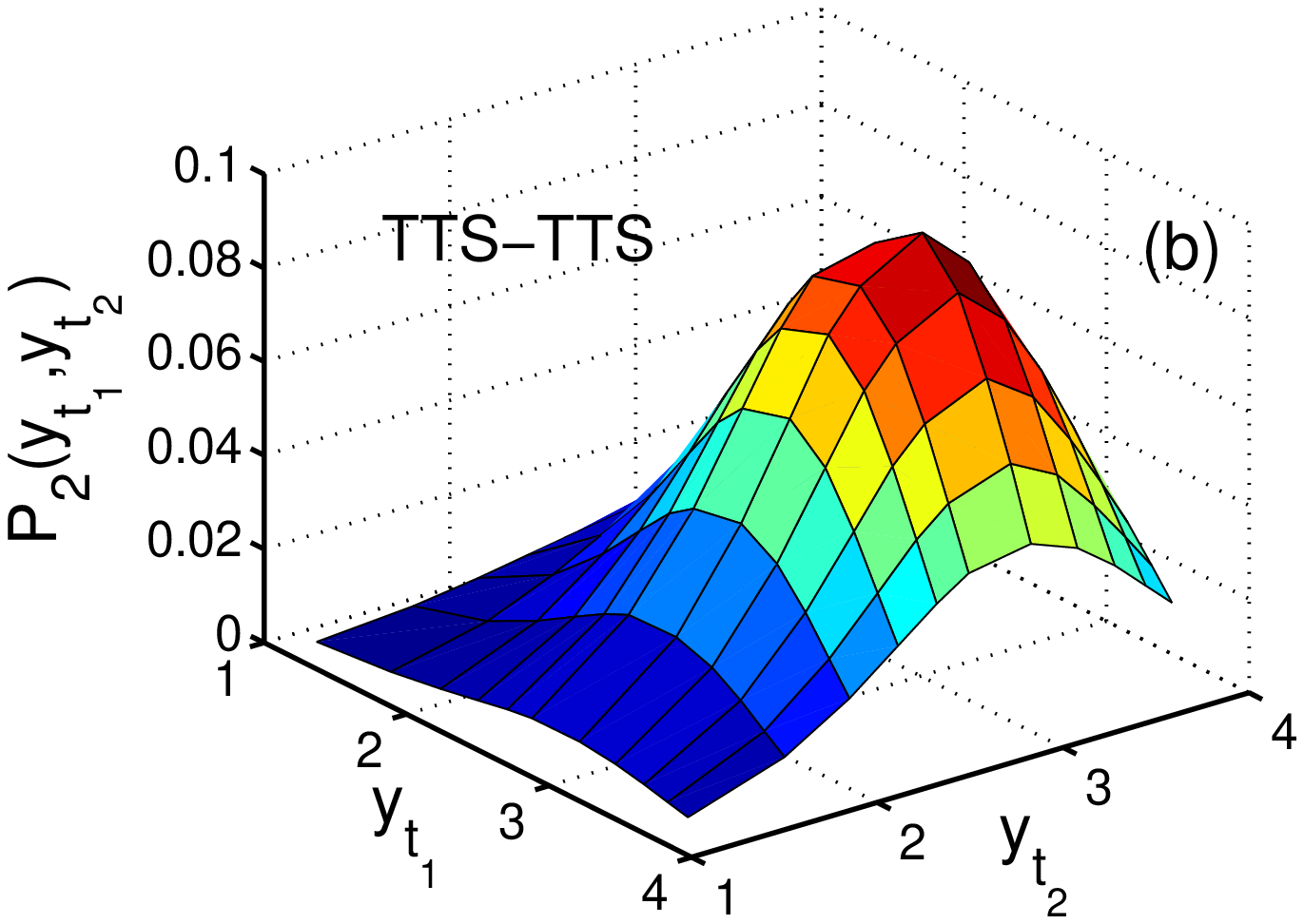}}
  \caption{(Color online) (TTS)(TTS) contribution at 2.76 TeV to $p$-$p$ correlation on $p_t$-$p_a$ and (b) $y_{t_1}$-$y_{t_2}$.}
\end{figure}

The above results are our best theoretical evidence for minijets. There are, of course, jets produced at all higher $p_T$. However, since inclusive two-particle distribution involves integration over the parton
 momenta $q_i$, higher-$p_T$ jets would not show up unless cuts in $p_T$ are made to select those jets. When that is done, one should be able to obtain
 peaks in the autocorrelation on the angular-difference variables $|\eta_1-\eta_2|$ and $|\phi_1-\phi_2|$, observed in the data \cite{lr,md,28.2} and
 calculated in  \cite{ch2}.

A summary of our findings in this section is that the peaking of two-particle correlations in $\pi\pi$ and $pp$ distributions provides indisputable evidence for minijets whose correlated shower partons are responsible for the phenomenon, as observed in \cite{lr}. Without particle identification, the correlation is due mainly to the process depicted in Fig.\ 3(a). Two-jet recombination is of negligible effect in correlations.

\section{Conclusion}

We have calculated the hadronic spectra for meson and baryon production in Pb-Pb collisions at LHC for $p_T$ in the range 0-16 GeV/c. Of particular concern in this work is the investigation of the extent to which multi-jet recombination is important in that $p_T$ range, since jet density is high at LHC. We have found that at 2.76 TeV the (SS)$^{2j}$ component in meson production makes negligible contribution at all $p_T$ compared to other components which are primarily TT and TS at low $p_T$, and then (SS)$^{1j}$ for $p_T>6$ GeV/c. For baryon production (SSS)$^{2j}$ is comparable to (SSS)$^{1j}$ at $p_T\approx 2$ GeV/c, so jet fragmentation alone in the conventional sense is not reliable at such low $p_T$. Recombination involving T is more important for $p_T<6$ GeV/c; in particular, (TSS)$^{nj}$ components with $n=1$ and 2 have comparable magnitudes, either one being larger than (SSS)$^{nj}$. Since that is the low-$p_T$ region  where the $B/M$ ratio is of the order of 1, the two-jet contributions to the inclusive spectra should not be ignored. The effect of 2-j recombination to the two-hadron correlation is, however, negligible.

It is not surprising that at 5.5 TeV the effect of multi-minijet contribution becomes more important. For proton production (TSS)$^{2j}$ is as large as (TSS)$^{1j}$ for $p_T<5$ GeV/c and (SSS)$^{2j}$ is comparable or larger than (SSS)$^{1j}$ for nearly all $p_T$. At $p_T\sim 6$ GeV/c all components except TTT and (SSS)$^{3j}$ are approximately similar in strength. Thus we predict that any calculation without 2-jet recombination would not be able to reproduce the data.

 We have shown by studying the single-particle distributions in $y_t$ that there are peaks at $y_t\approx 2.2$ and 2.8 for $\pi$ and $p$, respectively, verifying what has been observed experimentally by STAR \cite{tt,tt1,tt2}. Because of the definition in $y_t$ at low $y_t$, a Gaussian distribution in $y_t$ does not unambiguously indicate the existence of minijets. However, when two-particle correlation exhibits a broad peak in $y_t$-$y_t$ distribution, then the inference of minijets is inevitable. What we have shown is that the peak in $y_t$-$y_t$ receives its dominant contribution from (TS)(TS) in the case of $\pi\pi$ correlation, and from (TTS)(TTS) in $pp$ correlation. That is our explanation of the data on that peak observed by STAR \cite{lr}. The corresponding value of $p_T$ where the peak is located is around 1 GeV/c. Thus the minijets that give rise to the shower partons have a significant effect on low-$p_T$ physics through thermal-shower recombination.

The dominance of the role played by minijets in the low-$p_T$ region puts a new light on the subject of soft physics, since the conventional treatment by hydrodynamics does not take minijets into account explicitly. It raises the question on whether the non-flow component can ever be ignored at high collision energies, especially at LHC.
So far we have examined only $\phi$-averaged distribution in $p_T$ for central collisions. The problem that lies ahead is clearly the study of azimuthal anisotropy for non-central collisions. It is there that we must confront the LHC data on $v_n$ \cite{ka1}, and clarify the roles of minijets versus the fluctuations of initial-state configurations, both of which appear to have similar effects on the final-state hadronic observables.

\section*{Acknowledgment}

This work was supported,  in part,  by the U.\ S.\ Department of Energy under Grant No. DE-FG02-96ER40972 and by the Scientific Research Foundation
for Young Teachers, Sichuan University under No. 2010SCU11090 and
by the NSFC of China under Grant No.11147105.

 \begin{appendix}

  \section{Two-particle Correlations}

We summarize in this Appendix the formulas for the various terms contributing to $\pi$-$\pi$ correlation, followed by one term for $p$-$p$ correlation. With background subtraction defined for
$C_2(1,2)$ in Eq.\ (\ref{4.1}), it is sufficient to list only the non-factorizable terms involving S in various combinations.
The functions $\bar{F}_i(q,\kappa)$, ${\cal T}(p_1)$, $S_i(x)$, $\mathcal{S}^q(p_1,\kappa)$ and $\mathcal{S}^{qq}(p_1,p_2,\kappa)$ used
in following equations have been defined in Eqs.\ I-(38), I-(5), I-(B1), I-(A4) and I-(A12), respectively.

\subsection{(TS)(TS)}
\begin{equation}
\begin{array}{lll}
\dfrac{dN_{{\pi}{\pi}}^{(TS)(TS)}}{p_tp_adp_tdp_a}=\dfrac{1}{p_t^3p_a^3}\int\limits^{p_t}_0 dp_1\int\limits^{p_a}_0
dp_2\mathcal{T}\left(p_t-p_1\right)\mathcal{T}(p_a-p_2)\mathcal{S}^{qq}(p_1,p_2,\kappa).
\end{array}
\end{equation}

\subsection{(TS)(SS)$^{1j}$}
\begin{equation}
\begin{array}{lll}
\dfrac{dN_{{\pi}{\pi}}^{(TS)(SS)^{1j}}}{p_tp_adp_tdp_a}&=&\dfrac{1}{2}\left\{\dfrac{1}{p_t^3p_a}\int\limits^{p_t}_0 dp_1\mathcal{T}(p_t-p_1)\right.\\
&&\left.\times\sum\limits_i\int
\dfrac{dq}{q}\bar{F}_i\Big(q,\kappa\Big)\dfrac{1}{2}\left[S_i\left(\dfrac{p_1}{q}\right)\dfrac{1}{q-p_1}D_i^{\pi}\left(\dfrac{p_a}{q-p_1}\right)
+S_i\left(\dfrac{p_1}{q-p_a}\right)\dfrac{1}{q}D_i^{\pi}\left(\dfrac{p_a}{q}\right)\right] \right.\\
&+& \{p_t \leftrightarrow p_a\} \bigg\} .
\end{array}
\end{equation}

\subsection{(TS)(SS)$^{2j}$}
\begin{equation}
\begin{array}{lll}
\dfrac{dN_{{\pi}{\pi}}^{(TS)(SS)^{2j}}}{p_tp_adp_tdp_a} &=&\dfrac{\Gamma}{p_t^3p_a^3}\dfrac{1}{2}
\left[\int\limits^{p_t}_0 dp_1\mathcal{T}\big(p_t-p_1\big)\int\limits^{p_a}_0dp'_1
\mathcal{S}^{qq}(p_1,p'_1,\kappa)\mathcal{S}^q(p_a-p'_1,\kappa)
+ \{p_t \leftrightarrow p_a\} \right].
\end{array}
\end{equation}

\subsection{(SS)$^{1j}$(SS)$^{1j}$}
\begin{equation}
\begin{array}{lll}
\dfrac{dN_{{\pi}{\pi}}^{(SS)^{1j}(SS)^{1j}}}{p_tp_adp_tdp_a}&=&\dfrac{1}{p_tp_a}\sum\limits_i    \int_{p_t+p_a}
\dfrac{dq}{q^2}\bar{F}_i(q,\kappa)          \\
&&\times \dfrac{1}{2}\left[D_i^{\pi}\left(\dfrac{p_t}{q}\right)\dfrac{1}{q-p_t}D_i^{\pi}\left(\dfrac{p_a}{q-p_t}\right)
+   \{p_t \leftrightarrow p_a\}  \right].
\end{array}
\end{equation}

\subsection{(SS)$^{1j}$(SS)$^{2j}$}
\begin{equation}
\begin{array}{lll}
\dfrac{dN_{{\pi}{\pi}}^{(SS)^{1j}(SS)^{2j}}}{p_tp_adp_tdp_a}&=&\dfrac{\Gamma}{2}\left\{\dfrac{1}{p_tp_a^3} \int_0 ^{p_a}
dp_1\mathcal{S}^q(p_a-p_1,\kappa)\right. \sum\limits_i\int
\dfrac{dq}{q}\bar{F}_i(q,\kappa)    \\
&&\left.\times\dfrac{1}{2}\left[\dfrac{1}{q}D_i^{\pi}\left(\dfrac{p_t}{q}\right)S_i\left(\dfrac{p_1}{q-p_t}\right)
+\dfrac{1}{q-p_1}D_i^{\pi}\left(\dfrac{p_t}{q-p_1}\right)S_i\left(\dfrac{p_1}{q}\right)\right]\right.\\
&&+\  \{p_t \leftrightarrow p_a\}  \bigg\}.
\end{array}
\end{equation}

\subsection{(SS)$^{2j}$(SS)$^{2j}$}
\begin{equation}
\dfrac{dN_{{\pi}{\pi}}^{(SS)^{2j}(SS)^{2j}}}{p_tp_adp_tdp_a}=\dfrac{\Gamma}{p_t^3p_a^3}\int^{p_a}_0 dp'_1\int^{p_t}_0
dp_1 \mathcal{S}^{qq}(p_1,p'_1,\kappa) \mathcal{S}^{qq}(p_t-p_1,p_a-p'_1,\kappa).
\end{equation}

\subsection{(TTS)(TTS)}
\begin{eqnarray}
{dN_{pp}^{(TTS)(TTS)}\over p_tp_adp_tdp_a}&=& {g^2_{\rm st}N^2_pC^4\over m^p_T(p_t)m^p_T(p_a)p_t^{2\alpha+\beta+3}p_a^{2\alpha+\beta+3}} \int_0^{p_t} dp_1 \int_0^{p_t-p_1}
dp_2(p_1p_2)^{\alpha+1}(p_t-p_1-p_2)^{\beta}  \nonumber \\
&\times &e^{-(p_1+p_2)/T}
\int_0^{p_a} dp'_1 \int_0^{p_a-p'_1}
dp'_2(p'_1p'_2)^{\alpha+1}(p_a-p'_1-p'_2)^{\beta}e^{-(p'_1+p'_2)/T}  \nonumber \\
&\times &{\cal S}^{qq}(p_t-p_1-p_2,p_a-p'_1-p'_2,\kappa) .
\end{eqnarray}

\end{appendix}

\end{document}